\documentclass[12pt]{revtex4}
\usepackage[dvips]{graphicx}
\usepackage{delarray}
\usepackage{amsmath}
\usepackage{amssymb}
\usepackage{here}
\usepackage{color}
\usepackage{url}
\usepackage{tabularx,ragged2e,booktabs}
\newcommand{\ord}{\mathcal{O}}
\newcommand{\be}{\begin{equation}} \newcommand{\ee}{\end{equation}}

\newcommand{\vpa}{v_{\|}}
\newcommand{\vpe}{v_{\perp}}

\newcommand{\Tpa}{T_{\|}}
\newcommand{\Tpe}{T_{\perp}}

\newcommand{\p}{\partial}

\newcommand{\new}[1]{\textcolor{black}{#1}}
\newcommand{\gyro}{{\sc gyro}}
\newcommand{\neo}{{\sc neo}}
\newcommand{\transp}{{\sc transp}}
\newcommand{\Ev}{\mathbf{E}}
\newcommand{\Bv}{\mathbf{B}}
\newcommand{\rv}{\mathbf{r}}
\newcommand{\kpev}{\mathbf{k}_\perp}
\newcommand{\bv}{\mathbf{b}}
\newcommand{\vv}{\mathbf{v}}
\newcommand{\mN}{\mathcal{N}}
\newcommand{\mJ}{\mathcal{J}}
\newcommand{\energy}{\mathcal{E}}

\newcommand{\na}{\nabla}
\newcommand{\cd}{\cdot}
\newcommand{\al}{\alpha}
\newcommand{\tht}{\theta}
\newcommand{\ze}{\zeta}
\newcommand{\nar}{\nabla r}
\newcommand{\nabt}{\nabla \theta}
\newcommand{\naz}{\nabla \zeta}
\newcommand{\naa}{\nabla \alpha}
\newcommand{\nap}{\nabla \psi}
\newcommand{\kps}{k_\psi}
\newcommand{\kpe}{k_\perp}

\begin{document}

\begin{center}
\Large

{\bf Impurity transport in Alcator C-{Mod} in the presence of poloidal density variation induced by ion cyclotron resonance heating}\\
~\\*[0.5cm] \normalsize {A.~Moll\'en$^1$, I.~Pusztai$^{1,2}$,
  M.~L.~Reinke$^{2,3}$, Ye.~O.~Kazakov$^{4}$, 
  N.~T.~Howard$^{2}$, \new{E.~A.~Belli$^{5}$},
 T.  F\"ul\"op$^1$ 
 \new{and the Alcator C-Mod Team}%, S. Moradi$^1$
  \\
  \it\small $^1$ Department of Applied Physics,
  Chalmers University of Technology,
  G\"oteborg,
  Sweden}\\
{\it\small $^2$ Plasma Science and Fusion Center, Massachusetts
  Institute of Technology, Cambridge MA, USA }  \\
{\it\small $^3$ University of York, Heslington, York, United Kingdom}  \\
{\it\small $^4$ LPP-ERM/KMS, Association ``Euratom-Belgian State'', TEC
  Partner, Brussels, Belgium} \\ 
  {\new{\it\small $^5$ General Atomics, San Diego, CA, USA}} \\ \today
\end{center}
\begin{abstract}
  Impurity particle transport in an ion cyclotron resonance heated
  Alcator C-Mod discharge is studied with local gyrokinetic
  simulations and %an analytical 
  a theoretical 
  model including the effect of poloidal
  asymmetries and elongation.  In spite of the strong minority
  temperature anisotropy in the deep core region, the poloidal
  asymmetries are found to have a negligible effect on the \new{turbulent} impurity
  transport due to low magnetic shear in this region, in agreement
  with the experimental observations. According to %our 
  the theoretical
  model, in outer
  core regions poloidal asymmetries may contribute %considerably as to
  to the reduction of the impurity peaking,
  %reduce impurity transport, 
  but %effects driven by 
  uncertainties in
  atomic physics processes prevent quantitative comparison with
  experiments.

\end{abstract}

\maketitle

\section{Introduction}
Control of high-$Z$ impurity species is important for future magnetic
confinement fusion machines. The core concentration should be kept low
to avoid intolerable energy losses from radiation due to high-$Z$
elements and plasma dilution due to lighter impurities. Particularly in
the view of the next generation experiment ITER, which will employ 
a~tungsten divertor, with the inevitable presence of tungsten in the
plasma, this is an important issue.  Consequently the fusion community
is searching for means to control the impurities.

Core accumulation of impurities is in many cases connected to
neoclassical inward convection, in particular for toroidally rotating
plasmas \cite{AngioniEPS,ManticaEPS}, although turbulent transport due
to microinstabilities has also been shown to lead to peaked impurity
profiles. It has been experimentally observed that the core impurity
content can be reduced by applying Ion Cyclotron Resonance Heating
(ICRH) \cite{valisa,PutterichEPS}, however, the theoretical
understanding of this effect is not satisfactory.  In
Ref.~\cite{PutterichEPS}, where the application of central ICRH to a
JET plasma containing tungsten is investigated, it was suggested that
the reduced core content could be related to a change in
magnetohydrodynamic (MHD) activity.  The drive of this MHD-instability
and its effect on impurities have not yet been investigated.  Another
possibility is that the ICRH leads to a change in turbulence (from
ion-temperature-gradient (ITG) mode dominated to trapped-electron mode
(TEM) dominated), which affects the impurity transport.  In recent
studies it has been proposed that the change in radial impurity
transport observed under the application of ICRH, may occur due to the
impact the poloidal asymmetr{ies} ha{ve} on the radial turbulent
impurity transport
\cite{fulop,moradi2,albertICRH,albertTEM,varenna,neoVariation}. Poloidal
redistribution of the impurity species has indeed been measured when
the ICRH system was tuned to heat hydrogen minority ions in deuterium
plasmas \cite{reinke,MazonEPS}.  It is well known, that low-field-side
(LFS) ICRH %heating 
caused impurities to be poloidally asymmetrically
distributed and accumulated at the high-field-side (HFS) of the flux
surface.  A model for this effect was introduced in
Ref.~\cite{ingessonppcf} and further extended in Ref.~\cite{kazakov}.  It
was shown that ICRH forces minority ions to become more trapped on the
LFS of the torus due to the fact that ICRH increases mostly the
perpendicular energy of the resonant ions.  An excessive positive
charge due to the LFS accumulation of minority ions generates a
poloidally varying potential, which is too weak to affect the main
species, but can push impurities of high charge to the opposite side.
The purpose of the present work is to analyze if the gyrokinetic model
for the impurity peaking factor in the presence of ICRH-induced
poloidal asymmetries is consistent with experimental observations in
Alcator C-Mod.

Alcator C-Mod is a tokamak well-suited to study the effect of ICRH on
high-$Z$ impurities.  Intrinsic molybdenum impurity, originating from
plasma facing components is commonly observed in fractions of $n_z /
n_e = 10^{-4}$--$10^{-3}$. Furthermore molybdenum can be injected
using a laser blow-off technique that allows time dependent studies of
the impurity transport. High temperature anisotropies, $T_\perp/T_\|\sim 10$,
of the heated species can be generated due to the large deposited
power densities. The ion rotation speeds only reach around %$0.1 v_{Ti}$
\new{10\% of $v_{Ti}$}
(with $v_{Ti}=(2T_i/m_i)^{1/2}$ the ion thermal speed), since there is
no momentum input from neutral beams, which are not used for plasma
heating. Despite this, self-generated flows are sufficient to lead to
inertial effects for heavy impurities in some C-Mod plasmas. There are
various imaging and spectroscopic diagnostic systems on C-Mod, as
discussed in Sec.~\ref{sec:experimental}, which can be used to follow
the spatio-temporal evolution of the studied impurity. For our studies
we choose a discharge which is part of a scan dedicated to study the
effect of ICRH on high-$Z$ impurity transport, described more in
detail in Ref.~\cite{mattAPS}.

The remainder of the paper is organized as follows.  In
Sec.~\ref{sec:experimental} we describe experimental parameters for
the studied discharge and introduce the model for a poloidally varying
non-fluctuating potential under minority ICRH.  In
Sec.~\ref{sec:fluxes} we investigate linear stability characteristics
and nonlinear fluxes of the studied discharge.  We introduce the
gyrokinetic modeling of turbulent impurity transport in the presence
of poloidal asymmetries in Sec.~\ref{sec:impurity}, and apply it to
find the zero-flux impurity density gradient.  Finally the results are
discussed and summarized in Sec.~\ref{sec:conclusions}.

%%%%%%%%%%%%%%%%%%%%%%%%%%%%%%%%%%%%%%%%%%%%%%%%%%%%%%%%%%%%%%%%%%%%%%%%%%%%%%%%%%%%%%%%%%%%%%%%%%%%%%%%%%%%%%%%%%%%%%%%%%%%%%%%%%%%%%%%%%%%%%
%%%%%%%%%%%%%%%%%%%%%%%%%%%%%%%%%%%%%%%%%%%%%%%%%%%%%%%%%%%%%%%%%%%%%%%%%%%%%%%%%%%%%%%%%%%%%%%%%%%%%%%%%%%%%%%%%%%%%%%%%%%%%%%%%%%%%%%%%%%%%%

\section{Experimental discharge}\label{sec:experimental}
We will examine the $1.0-1.2\,\rm{s}$ time interval of the {Alcator
  C-Mod} discharge 1120913016, when hydrogen minority ICRH heating of
$3\,\rm{MW}$ was applied on the LFS at $r/a\sim 0.38$. Molybdenum,
apart from being an intrinsic impurity as the main material of the first
wall, was introduced in a controlled manner using a multi-pulse laser
blow-off system, while its dynamics %was 
\new{were} 
tracked by multiple radiation
imaging diagnostics \cite{mattAPS}.  Soft X-ray (SXR) and electron
cyclotron emission (ECE) based measurements give slightly different
values for the sawtooth inversion radius of the plasma, showing that
it was located in the region $0.32 < r/a < 0.40$. 
\new{An analysis of the SXR spectrograms indicates that no other MHD activity 
than the standard sawtooth instability was present.} 
\new{The average periods of the sawteeth instabilities were $\sim\!16~\mathrm{ms}$.} 
In our study we will 
assume $n_z / n_e = 5 \times 10^{-4}$, the exact concentration is not
critical for the results as long as impurities are in the trace
quantities in the sense that they do not dilute the plasma $Z n_z /
n_e \ll 1$ since then the turbulence is mostly unaffected by their
presence.  Molybdenum ($Z = 42$, $m_z = 95.96~\mathrm{u}$) is normally
not fully ionized, and we will assume a charge state
$\mathrm{Mo}^{+32}$, the Ne-like isoelectronic sequence, which is 
%expected to be one of 
the dominant charge %states 
\new{state} 
around mid-radius for
the range of %ion 
\new{electron} 
temperatures in this plasma 
\new{and is observed experimentally}.

Impurity measurements are taken during {the flat-top phases} of the
studied discharge, more than 200~ms after the ICRH has been switched
on.  The electron density and temperature profiles are measured using
Thomson scattering and electron cyclotron emission \cite{basse}.  The
main ion density is then calculated from the electron density, based
on $Z_{\mathrm{eff}}\sim 2.0$ calculated from neoclassical
conductivity and assuming constant impurity concentration profiles.
Main ion temperature and plasma rotation are calculated from measured
high resolution spectra of $\mathrm{Ar}^{+16}$ (argon) line emission,
and it is assumed that the main ion and impurity temperatures are
equal \cite{reinkeRSI}. The molybdenum density profiles are 
\new{also} 
reconstructed from 
\new{x-ray imaging} 
crystal spectroscopy data. 

The hydrogen minority fraction and the temperature anisotropy{, which
  define the strength of the arising electrostatic potential,} are
reconstructed using \transp~\cite{transp}.  The minority fraction is
almost constant at $f_H \equiv n_{H0} / n_{e0} = 0.05$, where the
index {`0'} refers to the flux-surface averaged density, throughout
the analyzed radial domain.  ICRH {power} is provided by three
antennas {operating at a slightly different frequencies}.  {The first
  antenna couples $P_1 = 0.75~\mathrm{MW}$ of ICRH power at a
  frequency $f_1 = 80.5~\mathrm{MHz}$, the second -- $P_2 =
  0.76~\mathrm{MW}$ at $f_2 = 80.0~\mathrm{MHz}$, and the third --
  $P_3 = 1.51~\mathrm{MW}$ at $f_3 = 78.0~\mathrm{MHz}$.}  The on-axis
toroidal magnetic field is $B_0 = 5.83~\mathrm{T}$, where the axis is
at $R_0 \approx 0.68~\mathrm{m}$. The plasma current is $I =
1.1~\mathrm{MA}$ and $q_{95} = 3.4$.

The normalized ICRH resonant magnetic field strength is estimated from
\cite{kazakov} $b_c = B_c/B_0 = 2\pi \left[m_H/\left(e
  B_0\right)\right] \cdot \left[\left(f_1 P_1 + f_2 P_2 + f_3 P_3
  \right) / \left(P_1 + P_2 + P_3 \right)\right] \approx {0.89}$, with
$B_c$ being the ICRH resonant magnetic field and $m_H$, $e$ are the
mass and charge of the hydrogen minority species.  The radial position
of the ICRH resonance is such that the resonant layer is tangential to
the flux surface at $\rho_c = {0.35}$ ($\rho =
\sqrt{\psi_t\left(r\right) / \psi_t\left(a\right)}$, where $2\pi
\psi_t$ is the toroidal flux), corresponding to ${r/a = 0.38}$, where
$r$ is the minor radius and $a$ the outermost minor radius 
\new{in the midplane}.

\begin{figure}[!ht]
  \scalebox{0.95}{%\includegraphics[width=0.33\textwidth]{figs/Alcator2012_MillerPoloidalAngle_DensityAndTemperatureProfileSmooth.eps}
  \includegraphics[width=0.33\textwidth]{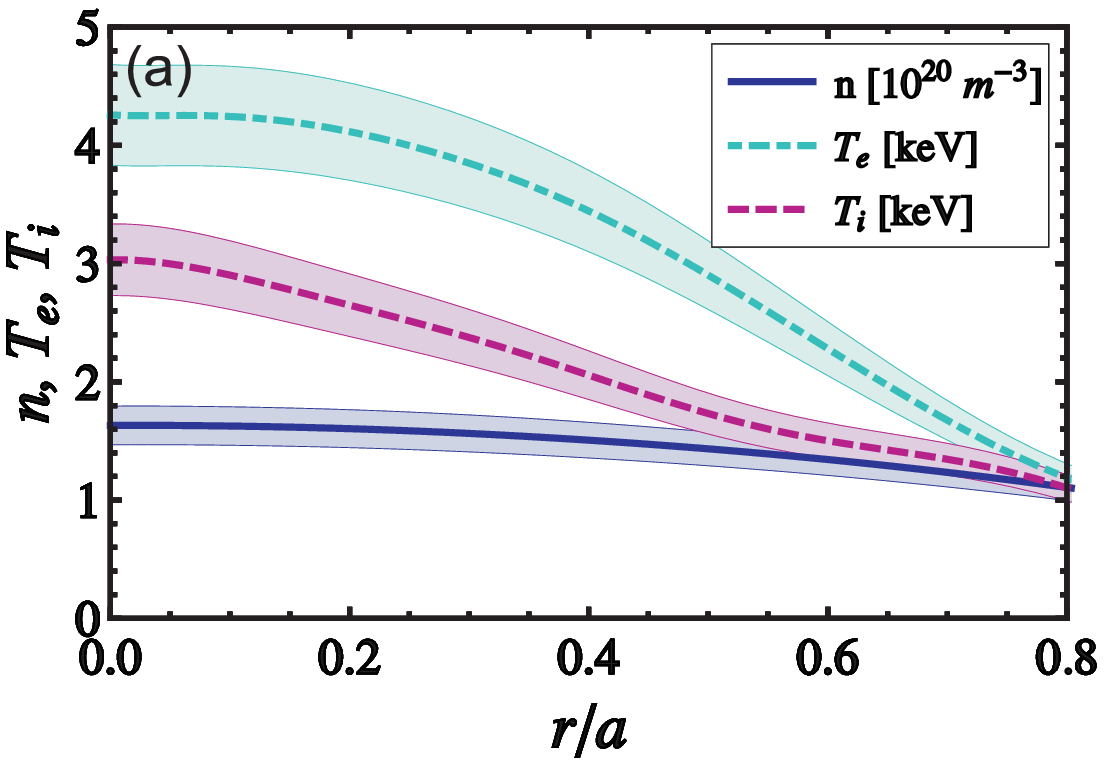}
 \includegraphics[width=0.33\textwidth]{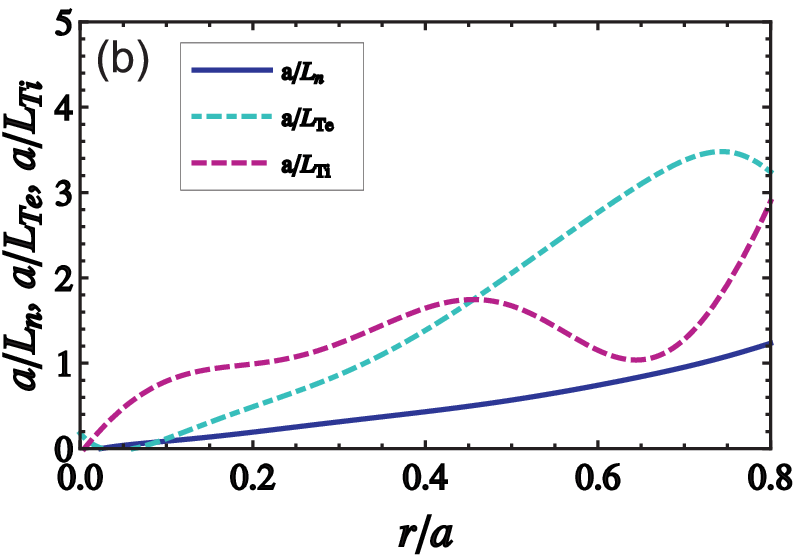}
 \includegraphics[width=0.33\textwidth]{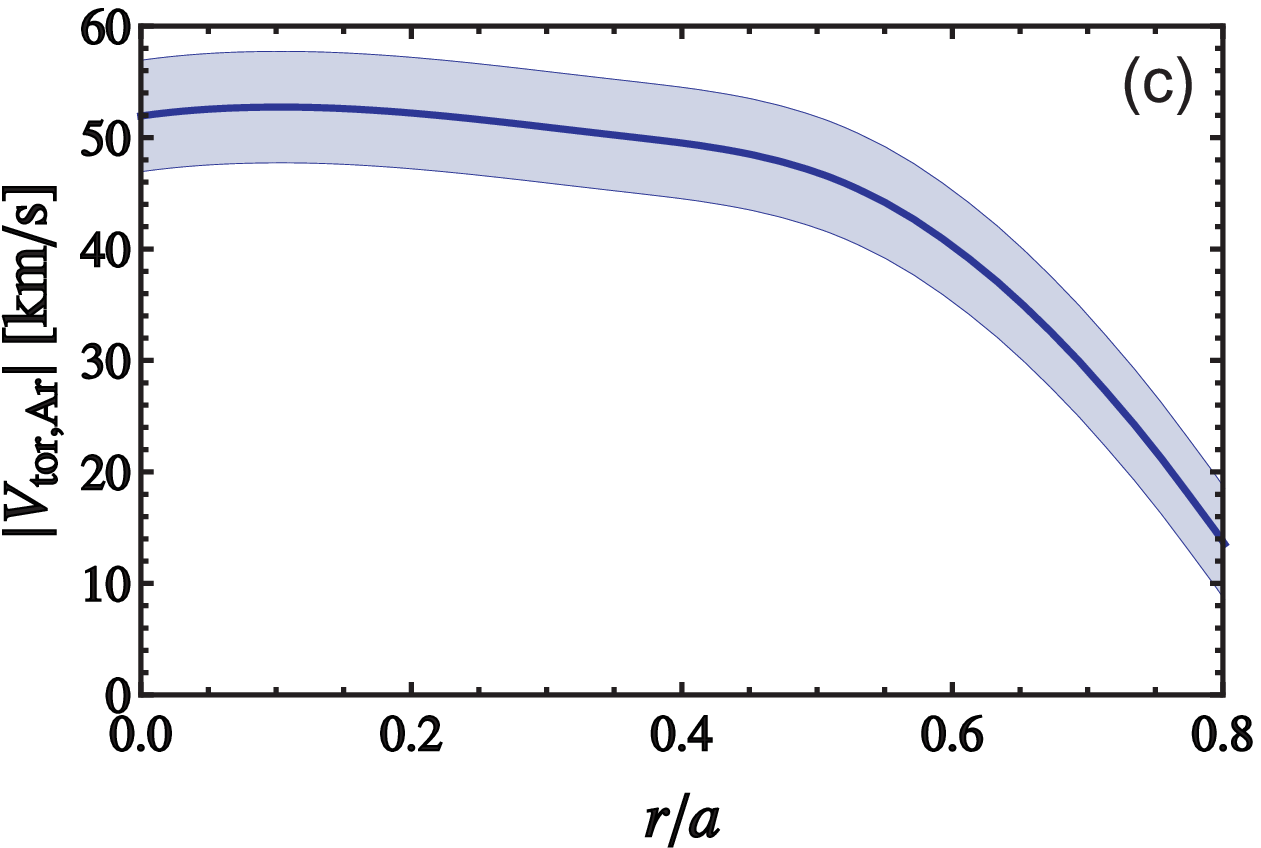}
 }
  \caption{(a)~Density and temperature ($n = n_e = n_i$, $T_e$, $T_i$)
  for the electrons and the main ions as functions of $r/a$, 
  with uncertainties represented by the shaded areas.
  (b)~Density- and temperature gradient scale lengths
($a/L_n$, $a/L_{Te}$, $a/L_{Ti}$) for the electrons and the main
ions as functions of $r/a$.
  (c)~Toroidal rotation of argon as function of $r/a$. 
  \new{Results are taken by averaging over $1.0 < t < 1.2$ in the studied discharge.}
  }
\label{fig:Profiles}
\end{figure}

Figure~\ref{fig:Profiles} shows experimental profiles of electron and
main ion densities ($n_e \approx n_i$, neglecting the much smaller
minority hydrogen and impurity concentrations) and temperatures
($T_e$, $T_i$), and their gradient scale lengths ($a/L_n$, $a/L_{Te}$,
$a/L_{Ti}$ defined by
$L_{n}=-\left[\partial\left(\ln{n}\right)/\partial r\right]^{-1}$ and
$L_{T}=-\left[\partial\left(\ln{T}\right)/\partial r\right]^{-1}$) for
the discharge, together with the argon rotation profile. 
The densities and temperatures are estimated to have an uncertainty of $\pm 10\%$. 
\new{Peak to peak changes in $T_e$ over time are approximately $1.5~\mathrm{keV}$, while the relative variations 
in $T_i$ and argon rotation are much smaller, and the changes in electron density are negligible.} 
The gradients are small close to the core region, but
larger at mid-radius and closer to the edge. We would thus expect to
find no unstable modes at low $r/a$ in our simulations. For $r/a
\lesssim 0.45$ we see that $a/L_{Ti} > a/L_{Te}$, while for $r/a
\gtrsim 0.45$ instead $a/L_{Ti} < a/L_{Te}$. It could hence be
expected that at lower radius the turbulence is ITG dominated, but
further out TEMs start to take over.

The radial domain we will focus on is $r/a = 0.20$--$0.60$, since that
is where the ICRH has a considerable impact on the minority
temperature anisotropy $\alpha_T \equiv \Tpe / \Tpa$ (see
Fig.~\ref{fig:VaryingPotential}(a)). Here, $\Tpe$ and $\Tpa$ are the
perpendicular and parallel temperatures of the minority species at the
studied flux surface.  In this radial domain the electron-ion
collision frequency 
%$\nu_{ei} = n_i e^4 \ln \Lambda/\left[4
%  \pi\epsilon_0^2 m_e^{1/2}(2T_e)^{3/2}\right]$ 
\new{
$\nu_{ei} = 4 \pi n_e e^4 \ln \Lambda/\left[m_e^{1/2}(2T_e)^{3/2}\right]$ 
(defined according to \gyro~conventions) 
}
  varies between
$\nu_{ei} = 0.02$--$0.05~c_s/a$, where $\ln \Lambda$ is the Coulomb
logarithm, $\epsilon_0$ is the vacuum permittivity, and $c_s =
\left(T_e/m_i\right)^{1/2}$ is the ion sound speed.  Furthermore, the
effective normalized electron pressure defined as $\beta_e = 8 \pi n_e
T_e / B^2$ varies between $\beta_e = 0.002$--$0.005$. Although it has been shown that electromagnetic
effects can be important for impurity transport even at low $\beta_e$
\cite{moradiEM}, these values are well below the critical value for
the onset of kinetic ballooning modes. We will thus restrict ourselves
to an electrostatic analysis.

\subsection{Poloidally varying equilibrium potential}
 
In Ref.~\cite{kazakov} it was discussed how ICRH affects the
poloidal minority ion distribution, and it was shown that when the
studied flux surface does not intersect the ICRH resonance, $B
\left(\theta\right) \geq B_c$, the minority density can be well
approximated with a sinusoidal poloidal variation.  In
Ref.~\cite{albertICRH} it was described how this could give rise to a
sinusoidal poloidal variation in the non-fluctuating potential over a
flux surface, given by
\begin{equation}
Ze\phi_E/T_{z}=-\mathcal{K} \cos \left(\theta-\theta_0\right),
\label{eq:Poloidal_potential}
\end{equation}
where $\mathcal{K}$ is the strength of the asymmetry and $\theta_0$
the poloidal location of the minimum of the minority distribution,
being $\theta_0 = \pi$ for ICRH.  The effect of the varying potential
on turbulent impurity transport has been studied in
Refs.~\cite{albertICRH,albertTEM,varenna,neoVariation}, and
expressions for the zero-flux impurity density gradient (impurity
peaking factor) have been presented.

\begin{figure}[!ht]
 \includegraphics[width=0.46\textwidth]{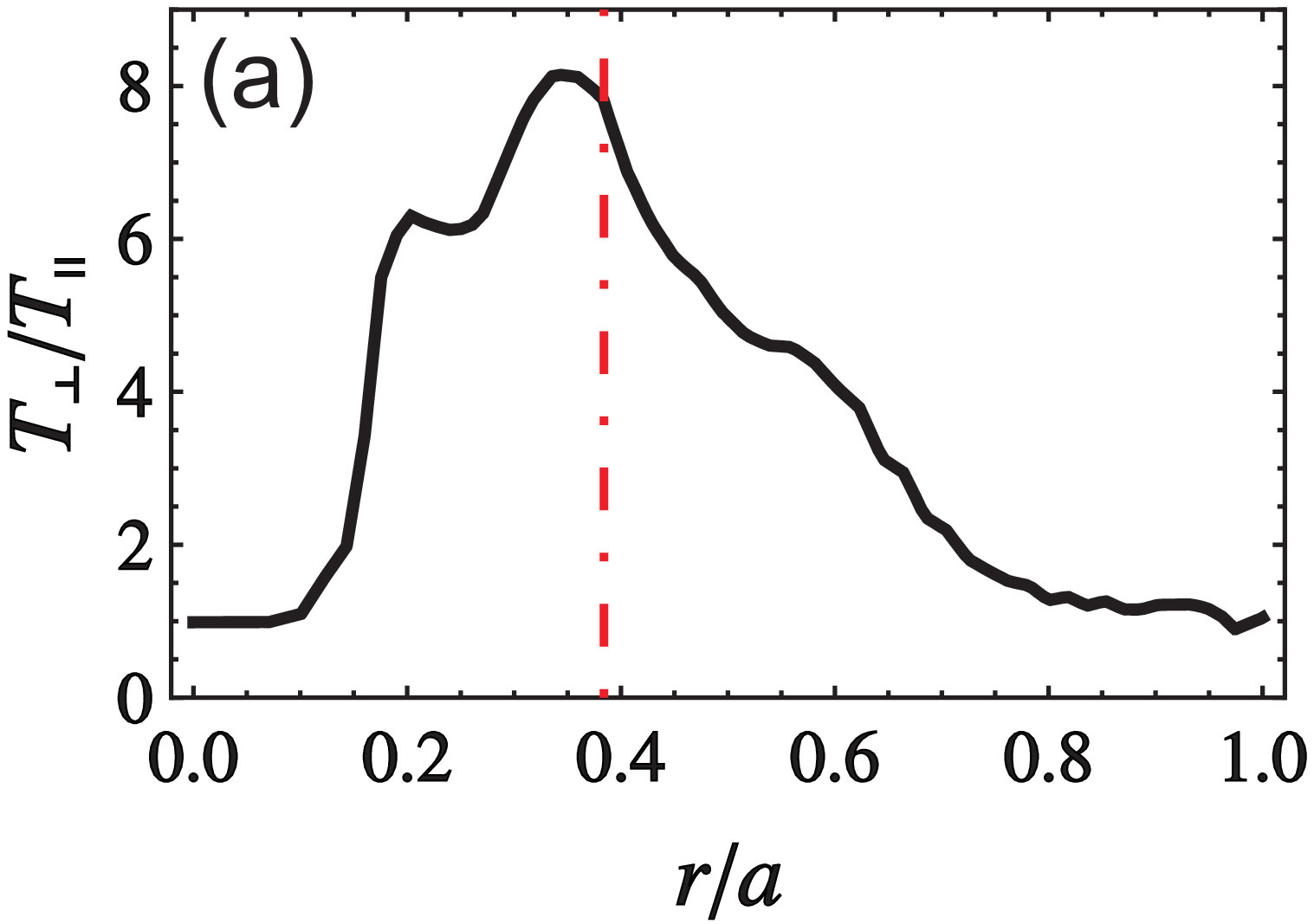}
  \includegraphics[width=0.53\textwidth]{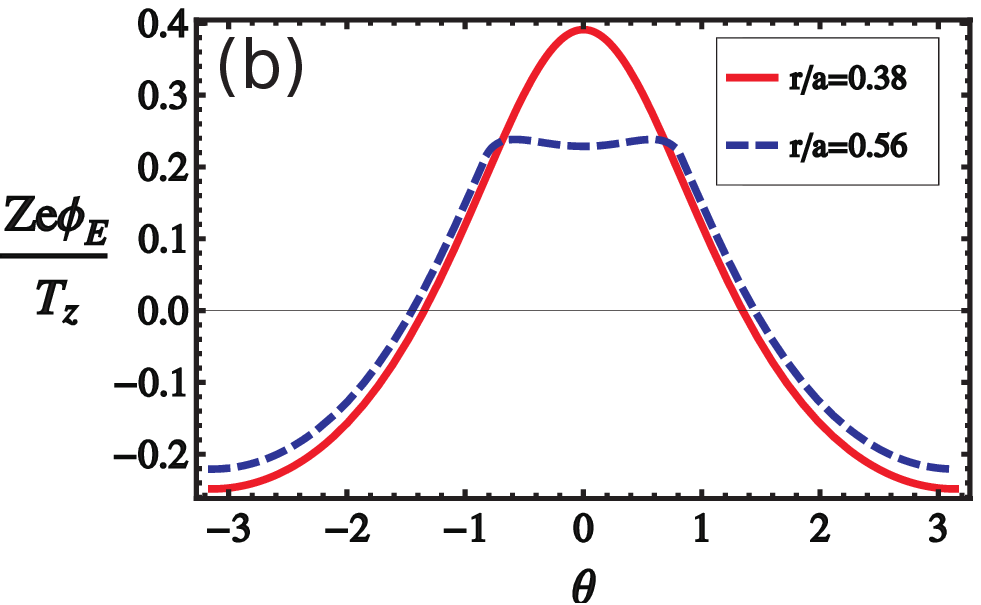}
  \caption{(a)~Temperature anisotropy of minority species $\alpha_T$ as
    function of radius {computed with \transp}. The location of
    the ICRH resonance is marked with a red dashed-dotted line.
    (b)~Non-fluctuating potential as function of poloidal angle at
    $r/a=0.38$ (red, solid) and $r/a=0.56$ (blue, dashed).
  }
\label{fig:VaryingPotential}
\end{figure}

In the present analysis we will not restrict ourselves to the
situation when the studied flux surface does not intersect the ICRH
resonance, but will also study radial locations where they do
intersect. For these flux surfaces a sinusoidal representation is not
a~good approximation for the varying potential and we will have to
resort to a numerical treatment.  To model the potential $\phi_E$ we
start with Eqs.~(2) and (3) in Ref.~\cite{kazakov}, describing how the
minority density varies over a flux surface
\begin{equation}
n_H = \left\{
\begin{array}{ll}
	 n_c \displaystyle\frac{T_{\perp a}}{T_{\perp}}, & B > B_c \\
	 n_c \displaystyle\left[\frac{T_{\perp a}}{T_{\perp}} +  \frac{T_{\perp b} - T_{\perp a}}{T_{\perp}} \left(\frac{T_{\perp}}{T_{\|}}\right)^{1/2} \left(\frac{B_c - B}{B_c}\right)^{1/2} \right], & B < B_c
\end{array}
\right.,
\label{eq:MinorityDensityGeneral}
\end{equation}
where
\begin{equation}
\begin{array}{l}
  T_{\perp a} = T_{\perp} \displaystyle\left[\frac{T_{\perp}}{T_{\|}} - \frac{B_c}{B} \left(\frac{T_{\perp}}{T_{\|}} - 1 \right)\right]^{-1} \\
  T_{\perp b} = T_{\perp} \displaystyle\left[- \frac{T_{\perp}}{T_{\|}} + \frac{B_c}{B} \left(\frac{T_{\perp}}{T_{\|}} + 1 \right)\right]^{-1}
\end{array}
\label{eq:Tperpab}
\end{equation}
and $n_c$ is the minority density on the flux surface at $B = B_c$.
The poloidal variation in Eqs.~{(\ref{eq:MinorityDensityGeneral})} and
{(\ref{eq:Tperpab})} enters through the magnetic field, which under
the assumption of large-aspect-ratio ($\epsilon \equiv r/R_0 \ll 1$)
is given by $ B = B_0/\left(1 + r \cos \theta/R_0 \right) $, where
$R_0$ is the major radius of the magnetic axis.  We define
$\tilde{n}_H \left(\theta\right) \equiv n_H \left(\theta\right) -
n_{H0}$ as the deviation from the average minority density over the
flux surface.  Assuming that the electrons, ions and impurities have a
Boltzmann response to the potential, $n_{\alpha} = n_{\alpha 0} \exp
(- e_{\alpha} \phi_E / T_{\alpha})$ and applying quasi-neutrality it
is found that the poloidally varying non-fluctuating potential is
given by
\begin{equation}
  \frac{Z e \phi_E}{T_z} = \frac{Z \tilde{n}_H / n_{e 0}}{{T_z}/{T_e}
    + n_{D0}/n_{e0}\left({T_{z}}/{T_{D}}\right)
+  \sum_{\rm imp} Z_{\rm imp}^2 \left({n_{\rm imp
        0}}/{n_{e 0}}\right) \left({T_{z}}/{T_{\rm imp}}\right) }
\label{eq:VaryingPotential}
\end{equation}
(note that this equation is similar to Eq.~(12) in \cite{ingessonppcf}
except that we have not assumed $n_{D 0} = n_{e 0}$ and $T_z = T_D$,
and that we allow for an arbitrary number of ion species).  When all
the ion temperatures are equal, 
%besides $T_{H}$ which is free to vary,
Eq.~(\ref{eq:VaryingPotential}) is reduced to $ Z e \phi_E/T_z = Z
(\tilde{n}_H /n_{e 0}) / \left({T_z}/{T_e} - f_H + Z_{\mathrm{eff}}
  \right) $, where $Z_{\mathrm{eff}} = \sum_i Z_i^2 n_i / n_e$ is the
effective ion charge.  In the case of low-field-side ICRH this
potential typically leads to an impurity distribution which is
displaced towards the inboard side of the flux surface.

Figure~\ref{fig:VaryingPotential} shows the radial profile of the
minority species temperature anisotropy, and also illustrates how $Z e
\phi_E/T_z$ from the model in {Eq.~(\ref{eq:VaryingPotential})} varies
with radius and poloidal angle.  Inside the ICRH resonance at {$r/a =
  0.38$}, the poloidal variation of $\phi_E$ exhibits a
sinusoidal-like behavior, but outside it gradually turns into a
variation with two peaks.  {These peaks correspond to the
  intersections of the flux surface with the almost vertical ICRH
  resonance layer, where the minority ions are accumulated.}  The
largest magnitude of $\phi_E$ almost coincides with the ICRH
resonance, and therefore we choose to analyze the radial location $r/a
= 0.38$ in our further study.  However at {this location} the magnetic
shear $s=(r/q)(dq/dr)$, where $q$ is the safety factor, is $s = 0.33$
which is a relatively low value. Earlier studies
\cite{albertICRH,albertTEM} have shown that due to its explicit
appearance in the $\Ev\times\Bv$-drift term, the magnetic shear can
have a strong impact on impurity peaking in the presence of poloidal
asymmetries, and in the case of low-field-side ICRH an increased shear
is expected to lead to a reduction of the peaking. The magnetic shear
increases with radius, and therefore we will also analyze a more outer
radial location $r/a = 0.56$ where $s = 0.78$. 
\new{
We note that from the calculated plasma rotation, 
we estimate the ratio of the $\mathrm{Mo}^{+32}$ rotation speed to its thermal speed to 
be around $v_{\mathrm{rot},z}/v_{Tz} \approx 0.8$ at both studied radial locations. 
}

\begin{figure}[!ht]
\includegraphics[width=0.49\textwidth]{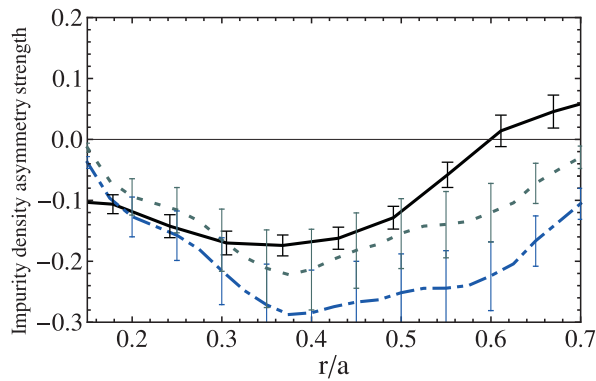}
  \caption{Comparison of the experimentally observed (solid) and
    numerically computed poloidal asymmetry due to ICRH only
    (dash-dotted) and with rotation retained (dotted).}
\label{fig:Asym}
\end{figure}

The poloidal variation of an impurity species in steady state is
approximately given by
\begin{equation}
n_z(\theta)\propto\exp\left\{-
\frac{Ze\phi_E}{T_z}-\frac{m_z\omega_z^2(R^2-R_0^2)}{2T_z}\left[1 -
  \frac{Z m_i}{m_z}\frac{Z_{\mathrm{eff}} T_e}{T_i + Z_{\mathrm{eff}}
    T_e}\right]\right\},
\label{eq:IncludeRot}
\end{equation}
where $\omega_z$ is the toroidal rotation frequency of the
impurities. The long wavelength electrostatic potential $\phi_E$
contains only the contribution from the poloidal variation of a
non-Boltzmann heated species, and it is given by
Eq.~(\ref{eq:VaryingPotential}). The second term in the square bracket
in Eq.~(\ref{eq:IncludeRot}) describes the effect of an electrostatic
potential set by the redistribution of all the other ion species on
the flux surface due to centrifugal forces. Figure~\ref{fig:Asym}
shows the impurity density asymmetry, that is the relative amplitude
of the $\cos \theta$ term in an expansion of the poloidal impurity
density variation in poloidal harmonics. The experimental value for
this quantity is inferred from emissivity measurements on the
mid-plane with AXUV, represented by the solid line with error bars,
described in Refs. \cite{reinke,reinkeThesis}. The dash-dotted line
shows the numerically computed asymmetry of the single charge state of
$\rm{Mo^{+32}}$, considering only the effect from the minority density
variation on the flux surface from ICRH and assuming
$Z_{\mathrm{eff}}=1.8$. In most of the plotted region the computed
in-out asymmetry is larger than the experimental. This discrepancy is
reduced considerably when rotation effects are retained; the dotted
curve is calculated keeping the $\omega_z^2$ part of
Eq.~(\ref{eq:IncludeRot}) assuming that the toroidal rotation speed is
similar to that measured for argon. Possible reasons for the remaining
disagreement might be uncertainties in the ion composition and in
rotation speed, or that the AXUV signal integrates several charge
states while the charge state composition varies radially due to the
temperature variation. Recent work shows that centrifugal and Coriolis
drift effects can have a non-negligible influence on radial impurity
transport \cite{Casson2010,Camenen2009,angioni2011,angioni2012},
however contributions from these drifts to the impurity peaking factor
are estimated to be negligibly small for the case studied here as
discussed later in Sec.~\ref{sec:impurity}. Therefore we will neglect
these effects together with the impact of rotation on the poloidal
distribution of particles and concentrate only on the impact of ICRH
driven asymmetries.

%%%%%%%%%%%%%%%%%%%%%%%%%%%%%%%%%%%%%%%%%%%%%%%%%%%%%%%%%%%%%%%%%%%%%%%%%%%%%%%%%%%%%%%%%%%%%%%%%%%%%%%%%%%%%%%%%%%%%%%%%%%%%%%%%%%%%%%%%%%%%%
%%%%%%%%%%%%%%%%%%%%%%%%%%%%%%%%%%%%%%%%%%%%%%%%%%%%%%%%%%%%%%%%%%%%%%%%%%%%%%%%%%%%%%%%%%%%%%%%%%%%%%%%%%%%%%%%%%%%%%%%%%%%%%%%%%%%%%%%%%%%%%

\section{Fluxes and mode characteristics}\label{sec:fluxes}
The characteristics of the turbulence present in the studied discharge
is analyzed with simulations using the gyrokinetic tool
\gyro~\cite{gyro}.  As discussed in Sec.~\ref{sec:experimental} we
neglect electromagnetic fluctuations, while collisions are included in
the simulations and modeled by the Lorentz operator
(pitch-angle-scattering).  If not otherwise mentioned only the
main ion species is included in the simulations, since the presence of
impurities in trace content does not affect the turbulence.

From linear gyrokinetic initial-value simulations we obtain the
perturbed electrostatic potential $\phi$ and the eigenvalues $\omega =
\omega_r + i \gamma$ for the most unstable mode, while any
sub-dominant modes are neglected.  The Miller parametrization of a model
Grad-Shafranov magnetic equilibrium is used, where the
$\ord\left(\epsilon\right)$ corrections to the drift frequencies are
retained.  We use flux-tube (periodic) boundary conditions, with a 192
point velocity space grid (8 energies, 12 pitch angles and two signs
of velocity), the number of radial grid points is 16, and the number
of poloidal grid points along particle orbits is 28 for trapped
particles.  The highest energy grid point is at $m_i
v^2/\left(2T_i\right)=6$.  Ions are taken to be gyrokinetic while the
electrons to be drift kinetic with the mass ratio
$\left(m_i/m_e\right)^{1/2}=60$.

\begin{figure}[!ht]
  \scalebox{0.95}{\includegraphics[width=0.49\textwidth]{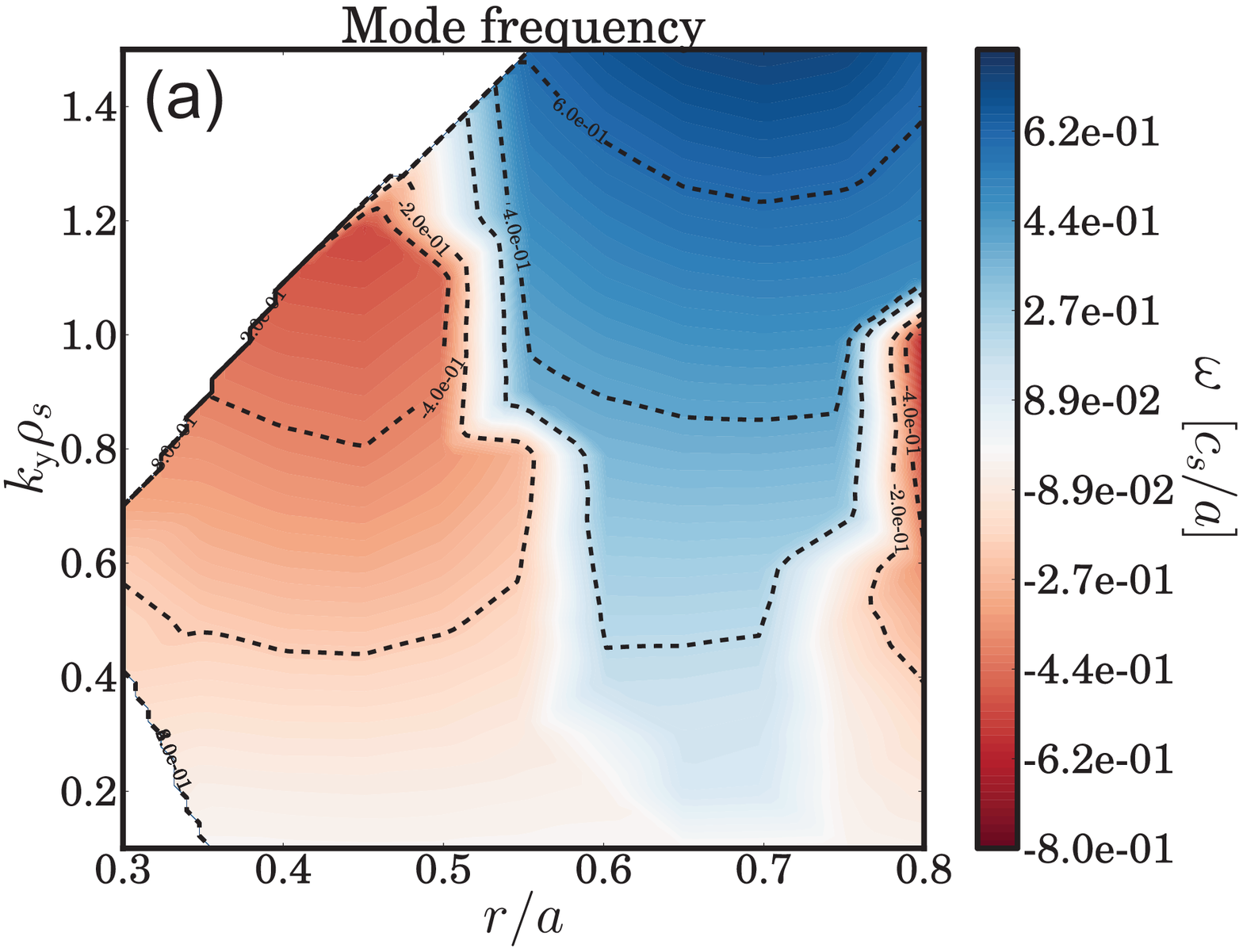}
    \includegraphics[width=0.49\textwidth]{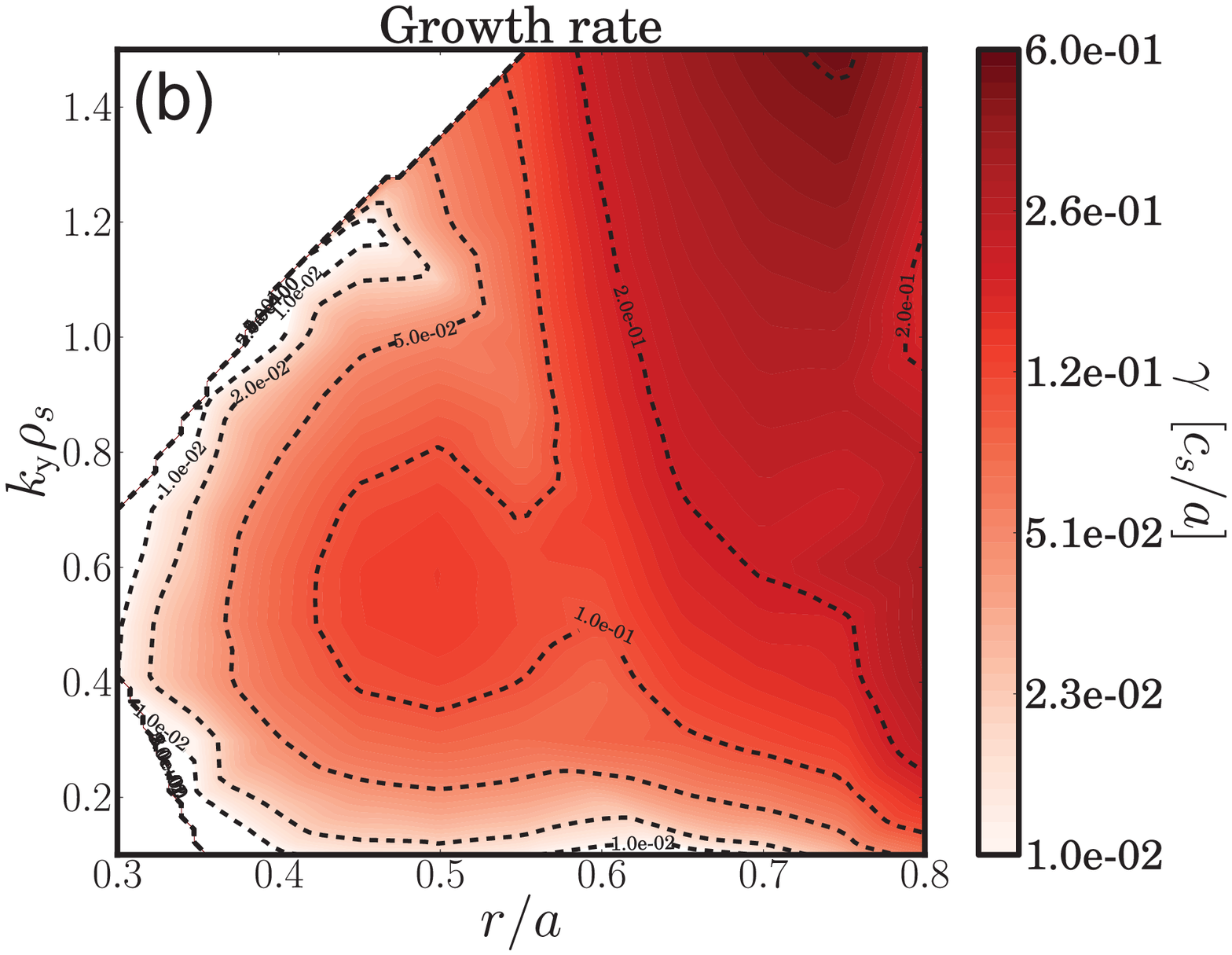}}
  \caption{Real mode frequency $\omega_r$ (a) and linear growth rate $\gamma$ (b) as functions of $r/a$ and $k_y \rho_s$ for the linearly most unstable mode, using the initial value solver in \gyro.}
\label{fig:Eigenvalues2DInitial}
\end{figure}

Figure~\ref{fig:Eigenvalues2DInitial} illustrates a map of linearly
unstable modes in $r/a$--$k_y \rho_s$ space (for $k_y \rho_s \leq
1.50$ by which the main part of the fluxes is normally driven), where
$k_y = n q /r$ is the binormal wave number with $n$ being the toroidal
mode number, and $\rho_{s0} = c_s/\Omega_{i0}$ is the ion sound Larmor
radius at $R_0$ with 
%$\Omega_{\alpha0}= e_{\alpha} B_0/m_{\alpha}$
\new{
$\Omega_{\alpha0}= e_{\alpha} B_0/m_{\alpha} c$
}
being the cyclotron frequency of species $\alpha$ at $B_0$.  There are
no linearly unstable modes for $r/a \lesssim 0.3$. In the vicinity of
the ICRH resonance and up to $r/a \lesssim 0.55$, the plasma is
clearly ITG dominated (ITG modes have negative real mode frequency
while TEMs have positive real mode frequency according to
\gyro~conventions).  For $r/a \gtrsim 0.55$ a TEM branch appears which
first is dominant only at higher wave numbers, but for $r/a \gtrsim
0.60$ it dominates the full wave number spectra.  However the ITG
branch continues to coexist as a sub-dominant branch, and becomes
dominant again for low wave numbers closer to the edge.  It is worth
noting that the linear growth rate of the ITG branch has a local peak
close to $r/a = 0.50$ and $k_y \rho_s = 0.60$, whereas the growth rate
of the TEM branch tends to increase with wave number. This mode is
likely to transit to an electron temperature gradient (ETG) mode at
even higher wave numbers.

\begin{figure}[!htbp]
  \scalebox{0.95}{\includegraphics[width=0.49\textwidth]{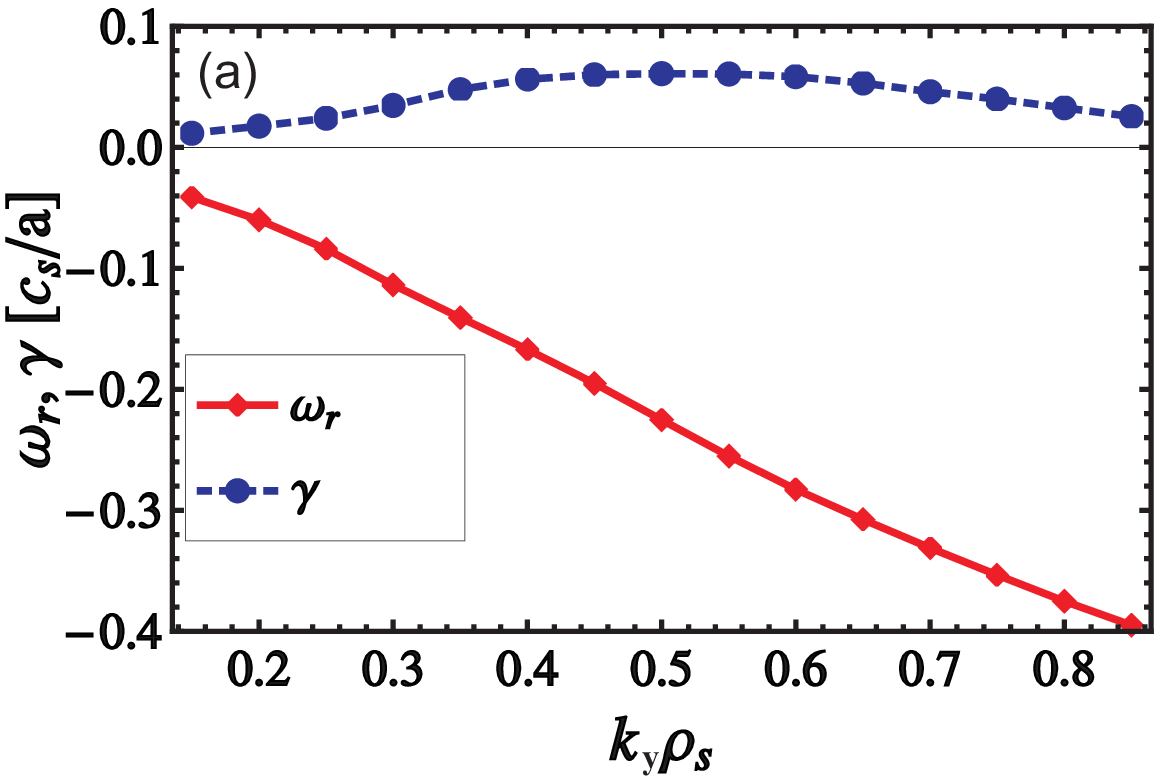}
    \includegraphics[width=0.49\textwidth]{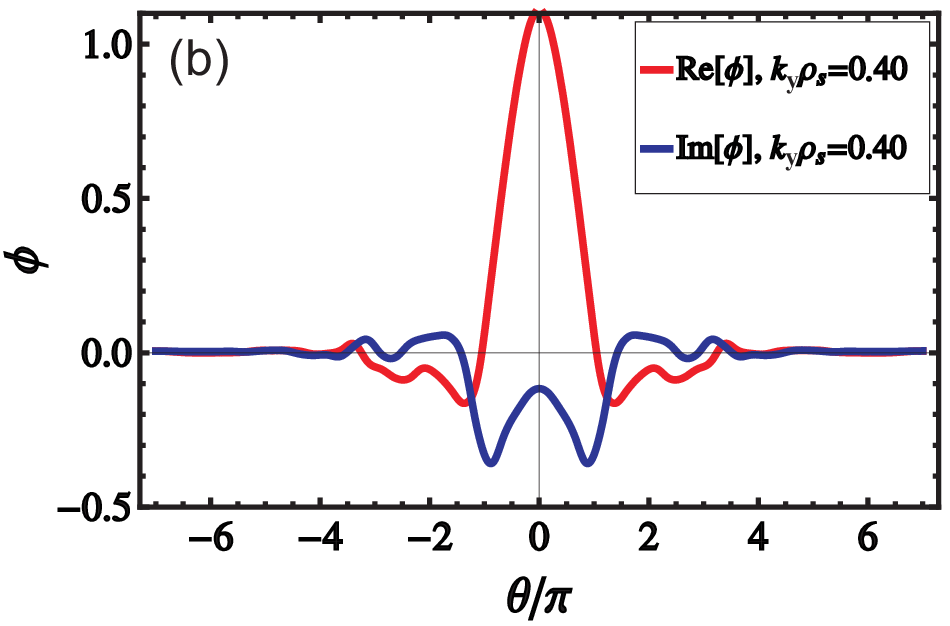}}
  \scalebox{0.95}{\includegraphics[width=0.49\textwidth]{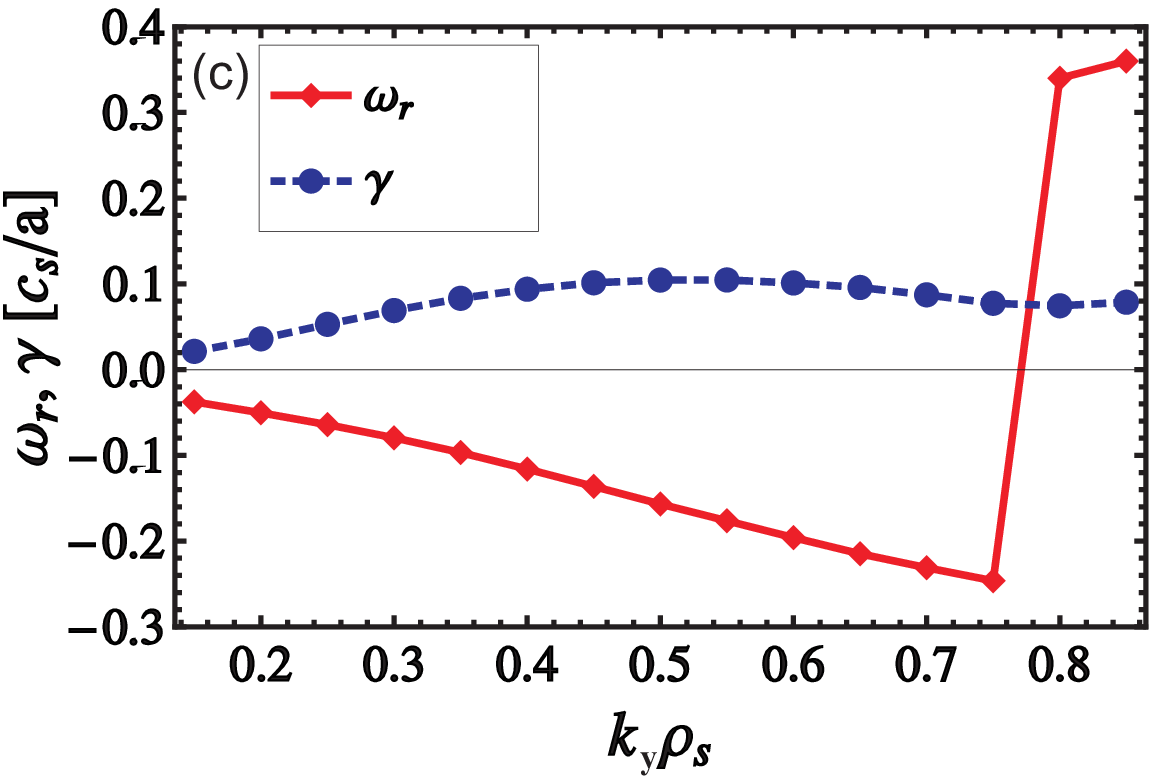}
    \includegraphics[width=0.49\textwidth]{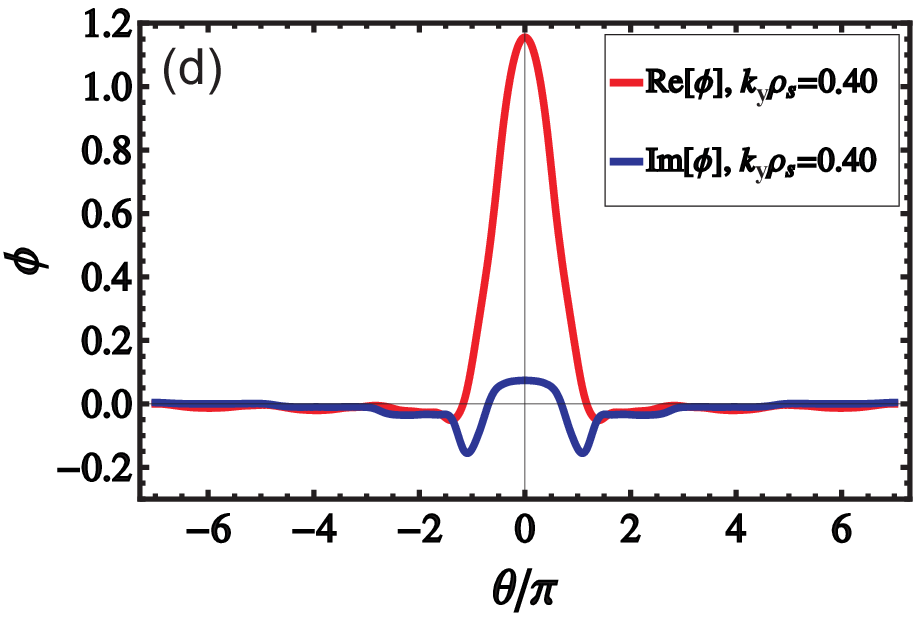}}
  \caption{(a,~c)~Real mode frequency $\omega_r$ (red solid line with diamonds) and linear growth rate $\gamma$ (blue dashed line with dots) as functions of $k_y \rho_s$ at $r/a = 0.38$ (a) and $r/a = 0.56$ (c).
  (b,~d)~Real (red line) and imaginary (blue line) part of the perturbed potential as functions of extended poloidal angle at $r/a = 0.38$ (b) and $r/a = 0.56$ (d) for $k_y \rho_s = 0.40$.}
\label{fig:EigenvaluesPotential}
\end{figure}

In Fig.~\ref{fig:EigenvaluesPotential} an excerpt of the eigenvalues
at $r/a = 0.38$ and $r/a = 0.56$ is shown, together with the linear
eigenfunction for the modes driving the largest fluxes nonlinearly,
$k_y \rho_s \approx 0.40$. Both radial locations are dominated by ITG
turbulence at lower wave numbers, and the eigenfunctions exhibit a
moderately ballooned structure concentrated to the interval $\theta \in
\left[-\pi, \pi\right]$.

\begin{figure}[htbp]
  \scalebox{0.95}{\includegraphics[width=0.49\textwidth]{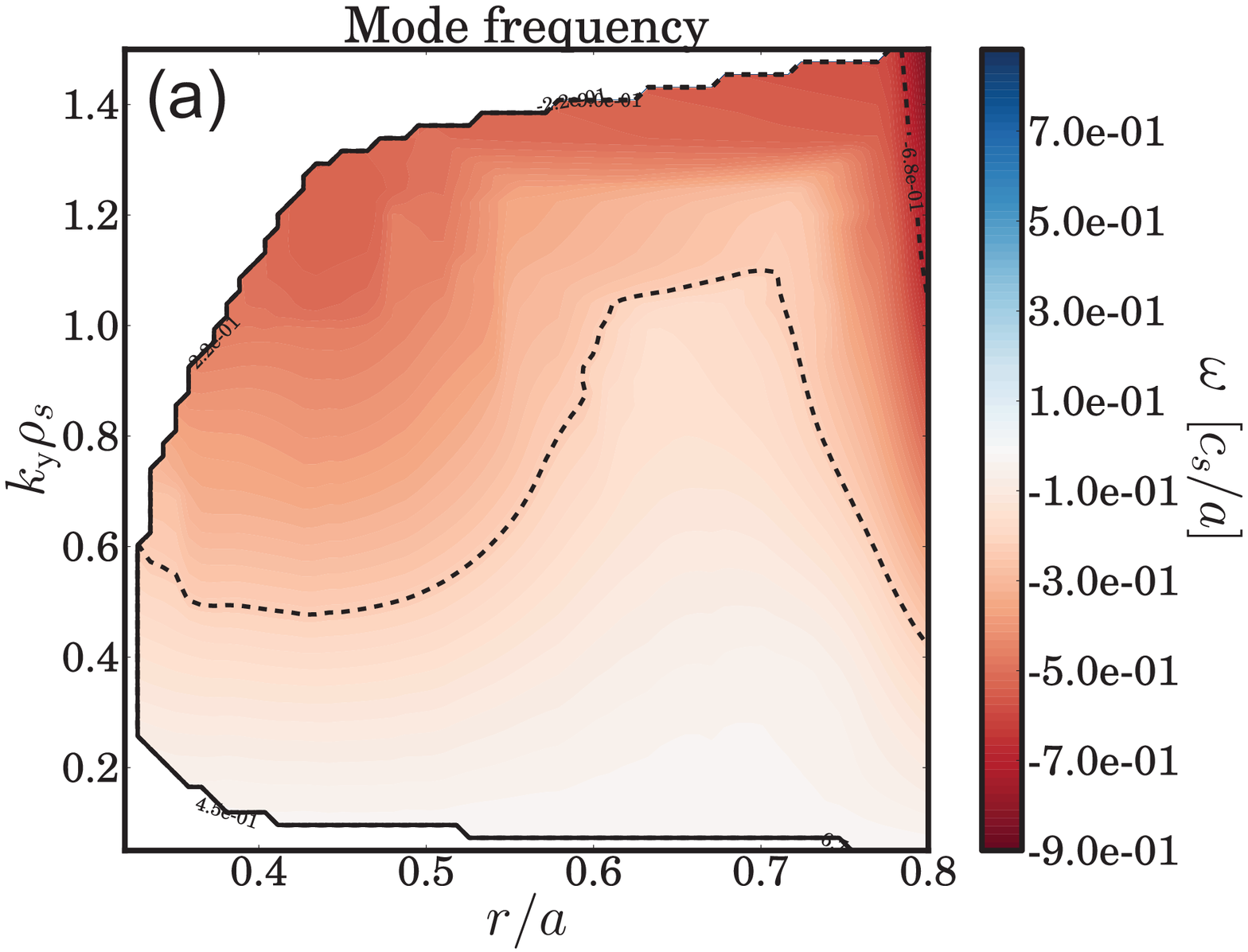}
    \includegraphics[width=0.49\textwidth]{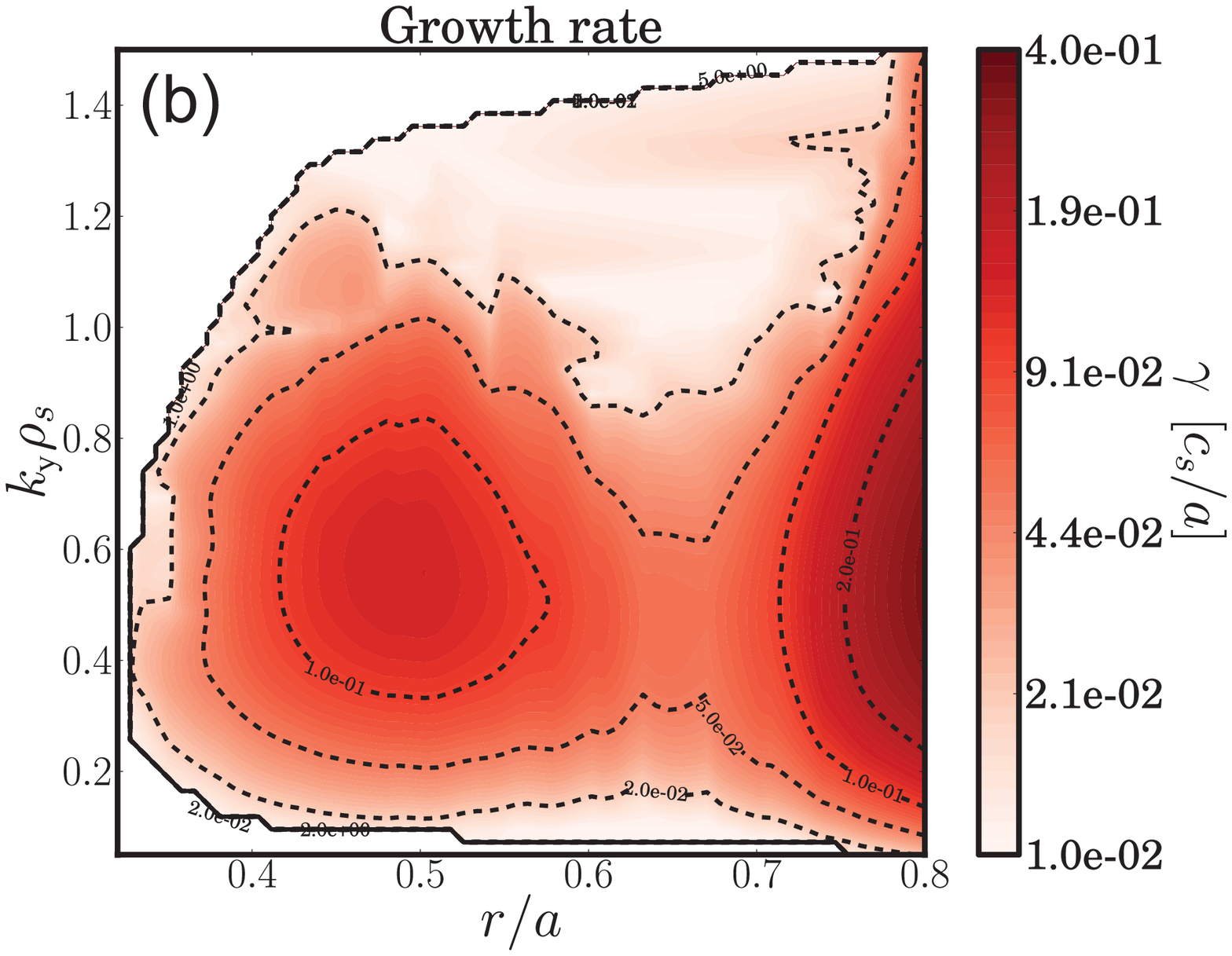}}
  \scalebox{0.95}{\includegraphics[width=0.49\textwidth]{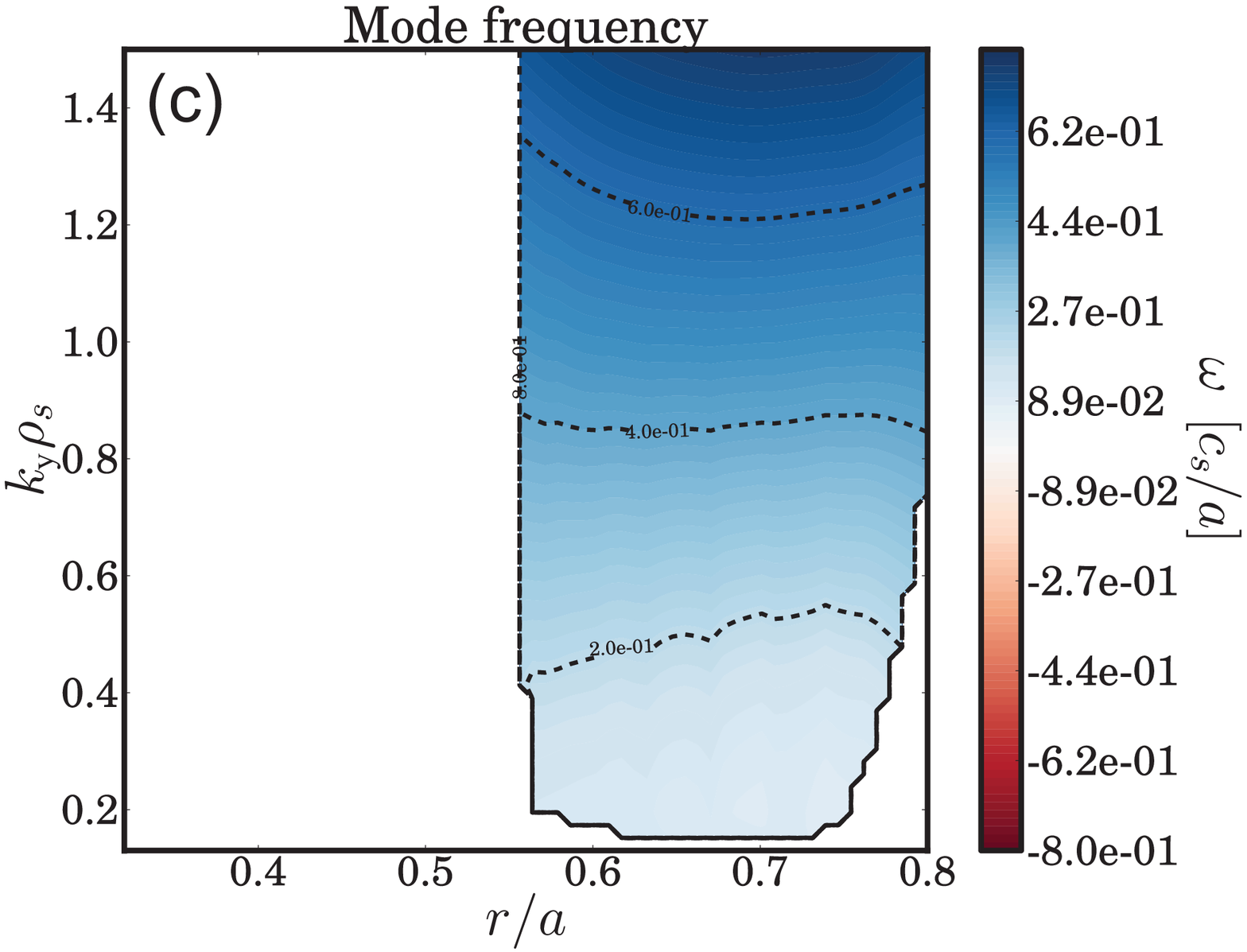}
    \includegraphics[width=0.49\textwidth]{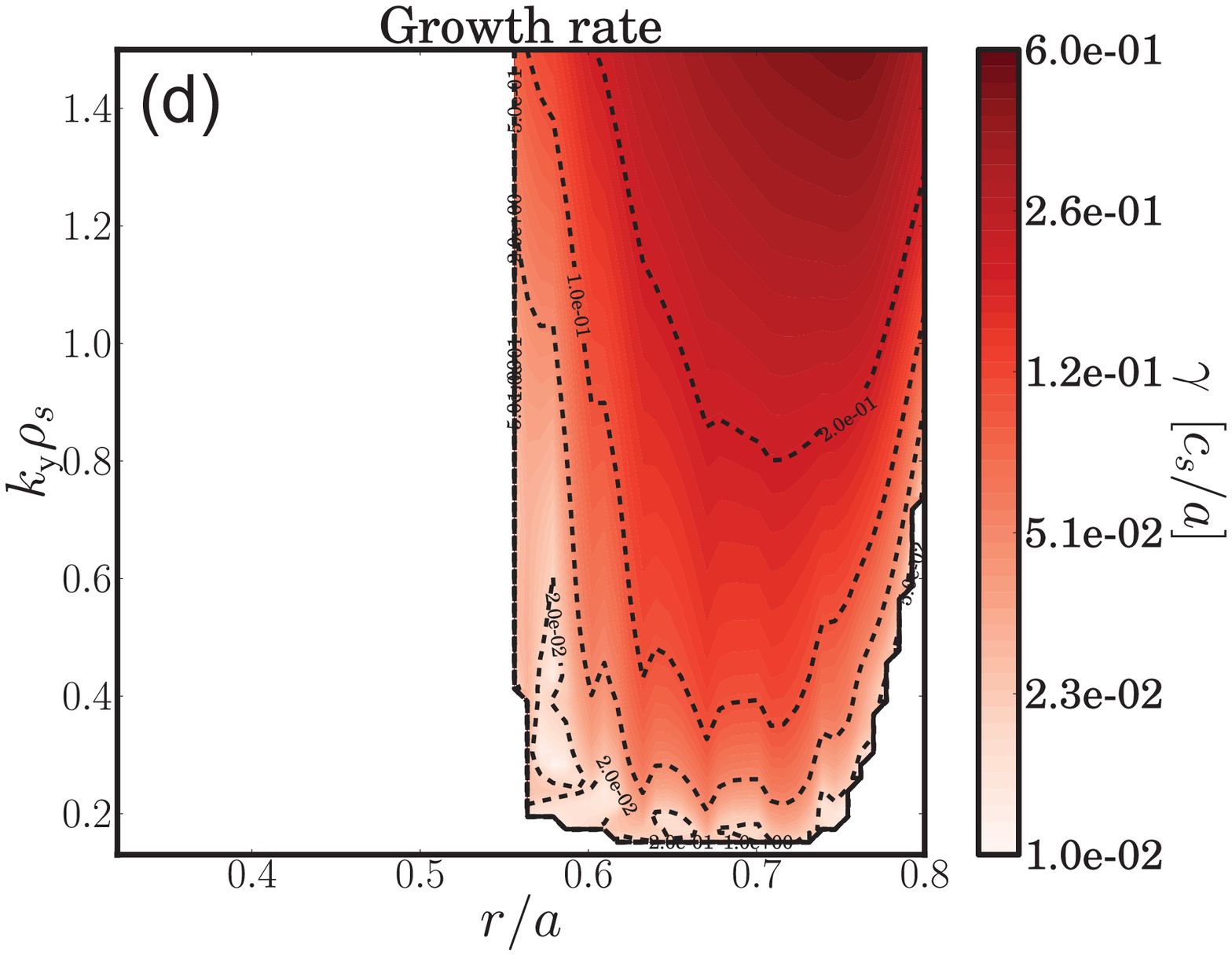}}
  \caption{Real mode frequency $\omega_r$ (a, c) and linear growth rate $\gamma$ (b, d) as functions of $r/a$ and $k_y \rho_s$, following the ITG branch (a, b) and the TEM branch (c, d) with the eigenvalue solver in \gyro.}
\label{fig:Eigenvalues2D}
\end{figure}

To get a wider picture, we trace the two branches observed in
Fig.~\ref{fig:Eigenvalues2DInitial} (i.e. the ITG and the TEM branch)
using the eigenvalue solver in \gyro. The results are illustrated in
Fig.~\ref{fig:Eigenvalues2D}. 
We see that the ITG branch exists in the full radial domain where
turbulent modes are found (i.e. for $r/a \gtrsim 0.3$) although it is
sub-dominant to the TEM branch for $0.55 \lesssim r/a \lesssim
0.75$. The branch is limited to $k_y \rho_s \lesssim 1.40$, and the
largest linear growth rates are found for $k_y \rho_s \approx 0.50$.
The TEM branch only exists at $r/a \gtrsim 0.55$. Both its growth rate
and linear mode frequency typically increase with increasing wave
number.

\begin{figure}[htbp]
\centering
\begin{minipage}{.5\textwidth}
 \includegraphics[height=0.6\textwidth]{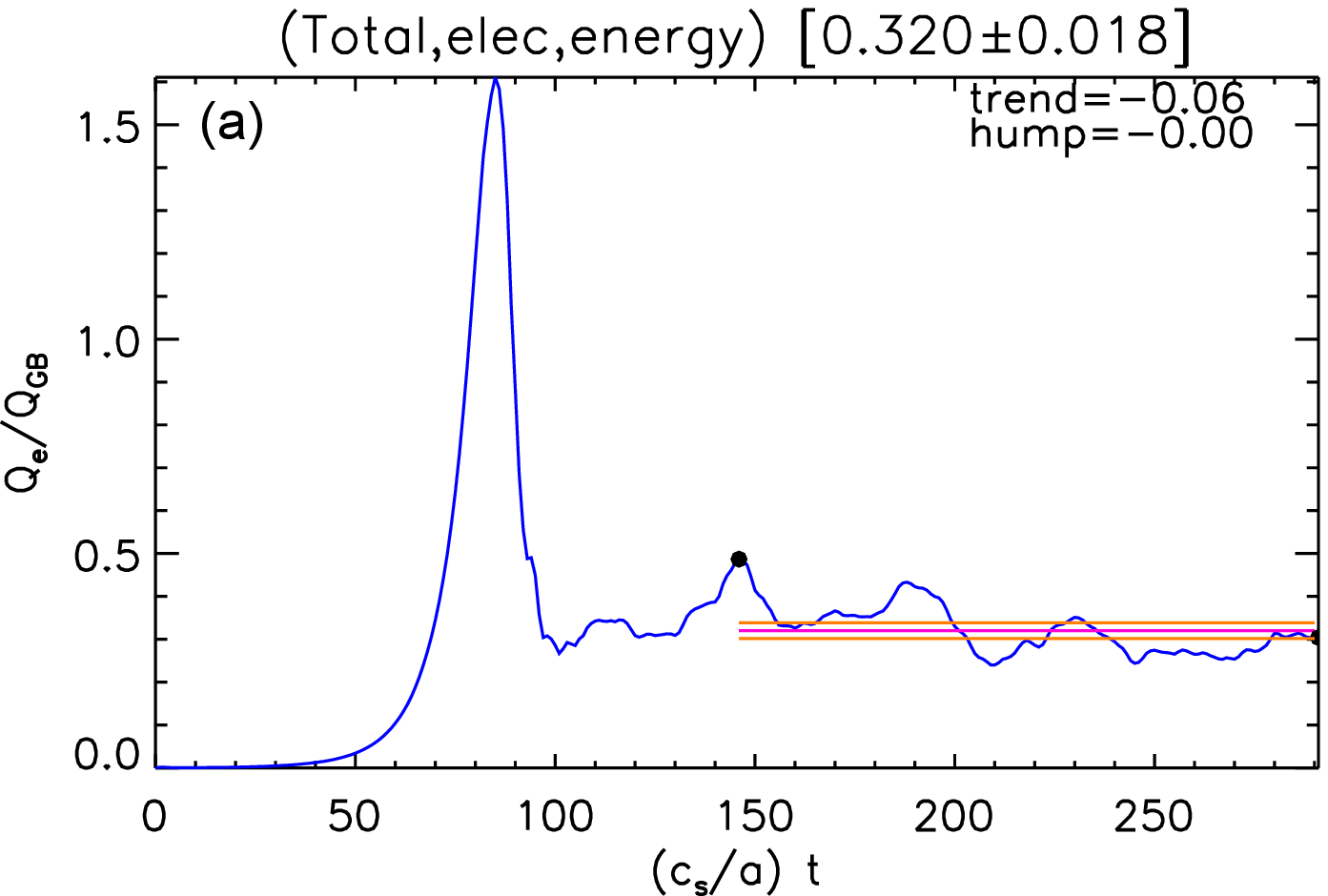}
\hbox{\hspace{0.5em}\includegraphics[height=0.6\textwidth]{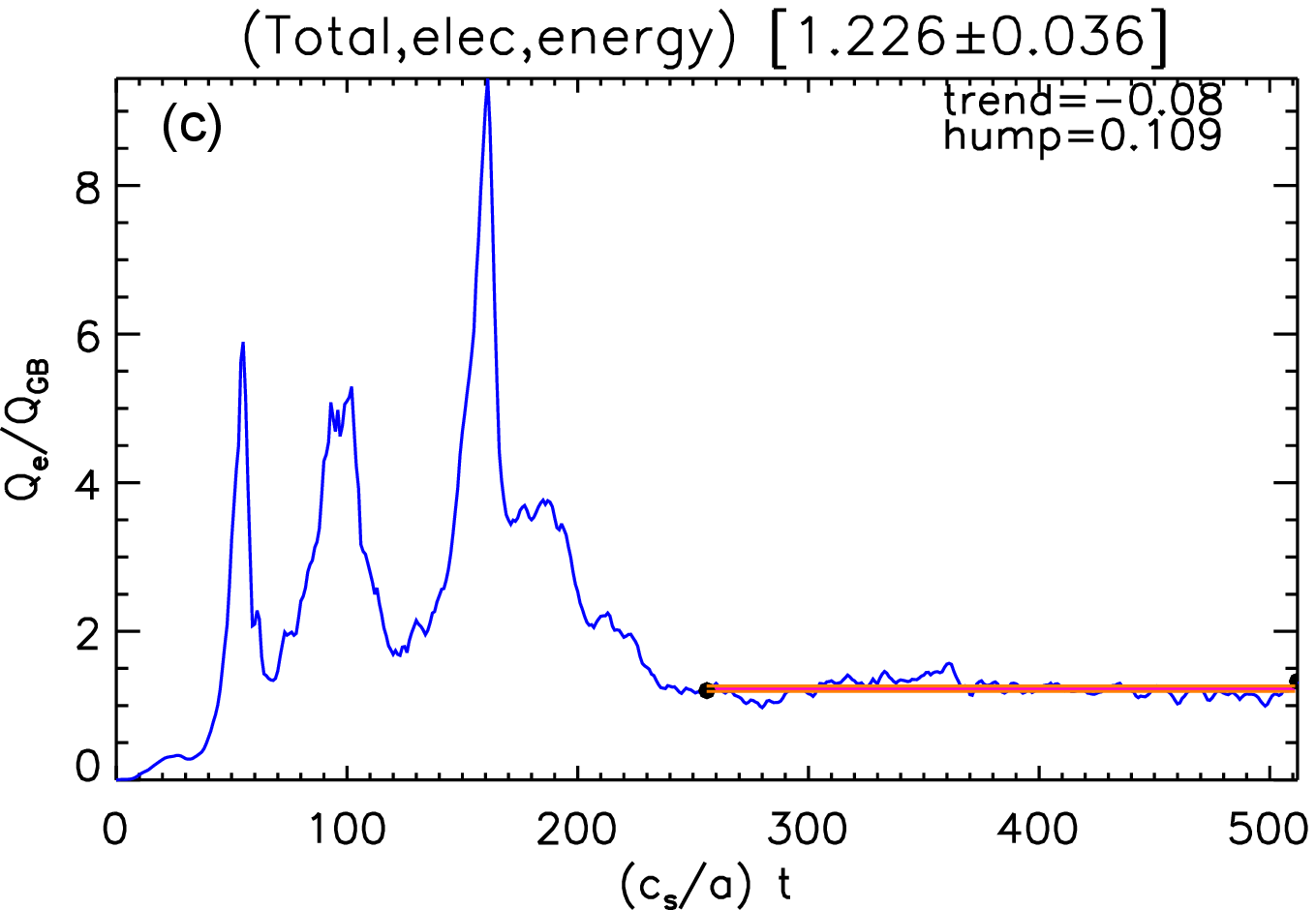}}
\end{minipage}%
\begin{minipage}{.5\textwidth}
\includegraphics[height=0.6\textwidth]{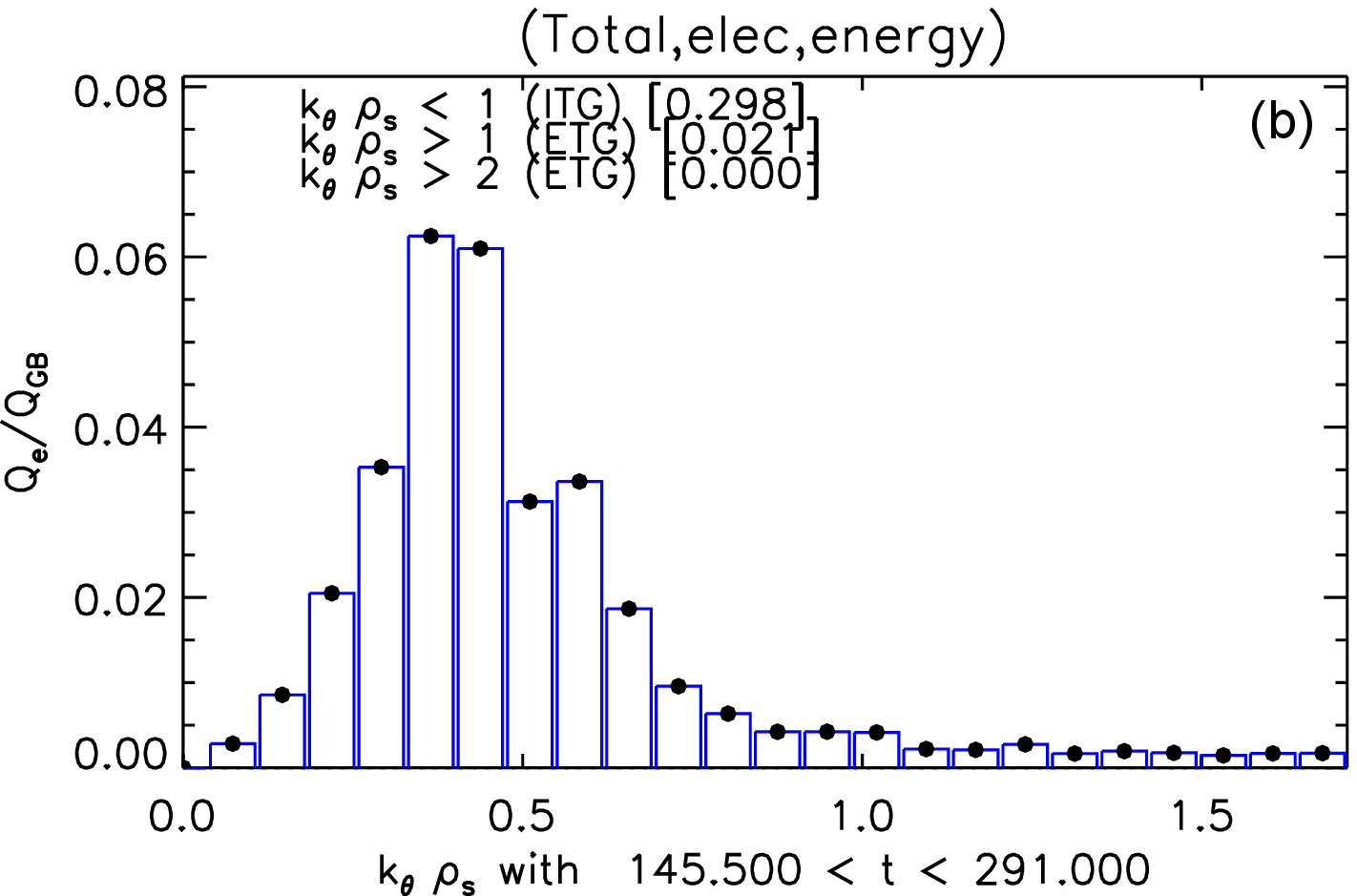}
 \includegraphics[height=0.6\textwidth]{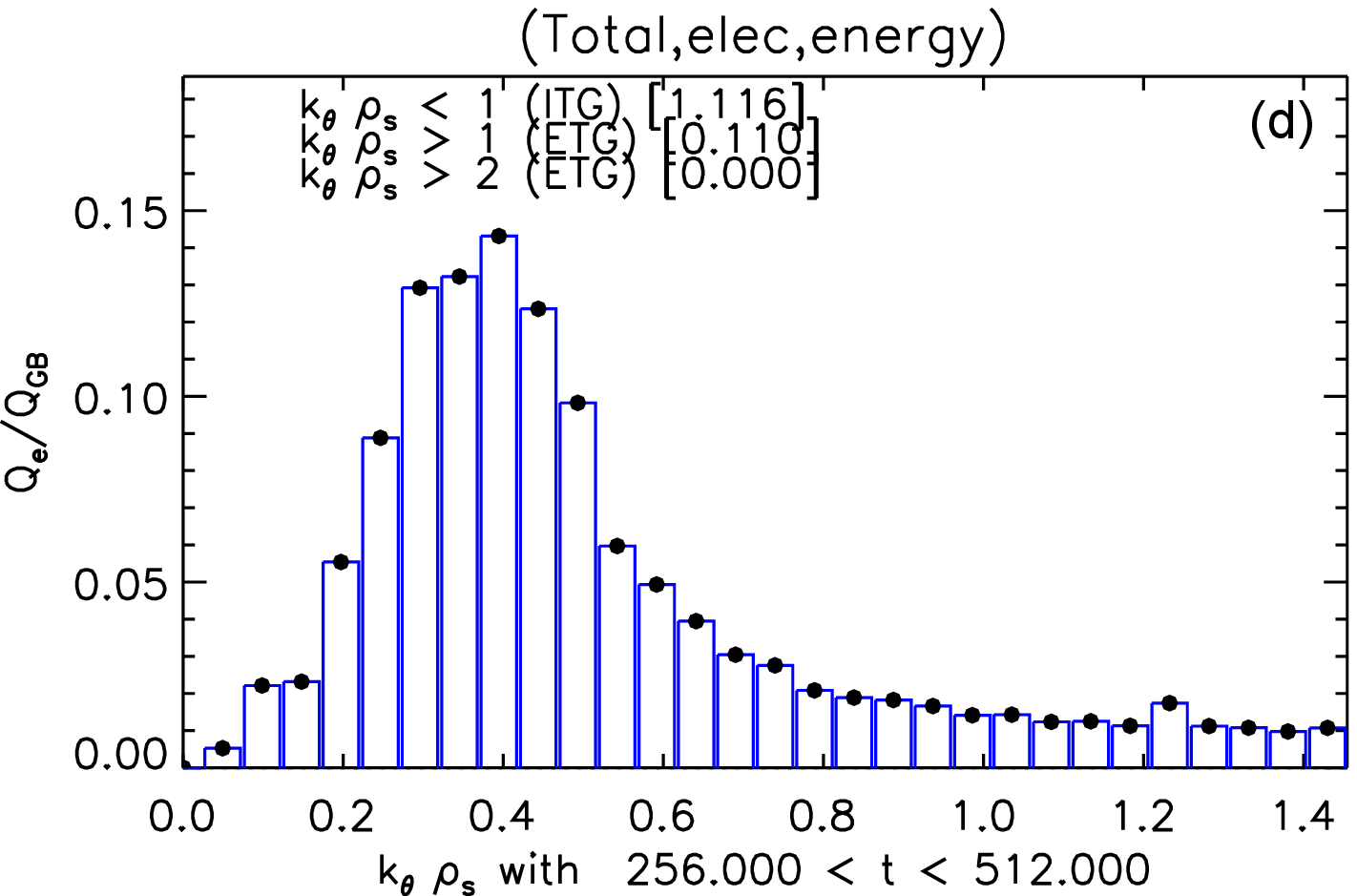}
\end{minipage}%
    \caption{Electron energy fluxes from \gyro~simulations at $r/a=0.38$
    (a,~b) and $r/a=0.56$ (c,~d).  (a,~c)~Time-evolution of the
    fluxes. (b,~d)~Poloidal wave number spectra from the time when the
    simulations have reached a saturated state.}
\label{fig:NonlinearGYRO}
\end{figure}

The nonlinear electrostatic \gyro~simulations are also performed with
gyrokinetic ions and drift kinetic electrons and use the same velocity
resolution as the linear simulations.  300 radial grid points are
used, and at least 24 toroidal modes to model ${1/5}^{\mathrm{th}}$ of
the torus, with the highest resolved poloidal wave number being
$k_y\rho_s \approx 1.5$.  The simulations are run with the integration
time step $\Delta t = 0.01 a/c_s$ for $t > 150 a/c_s$, when the fluxes
have saturated.

Figure~\ref{fig:NonlinearGYRO} shows how the electron energy fluxes
evolve over time and the poloidal wave number spectra for both
$r/a=0.38$ and $r/a=0.56$. The main part of the fluxes are clearly
driven in the regime around $k_y \rho_s = 0.40$, which at both radial
locations is ITG dominated (see Fig.~\ref{fig:EigenvaluesPotential}).

\begin{figure}[htbp]
\begin{minipage}{.5\textwidth}
\includegraphics[height=0.58\textwidth]{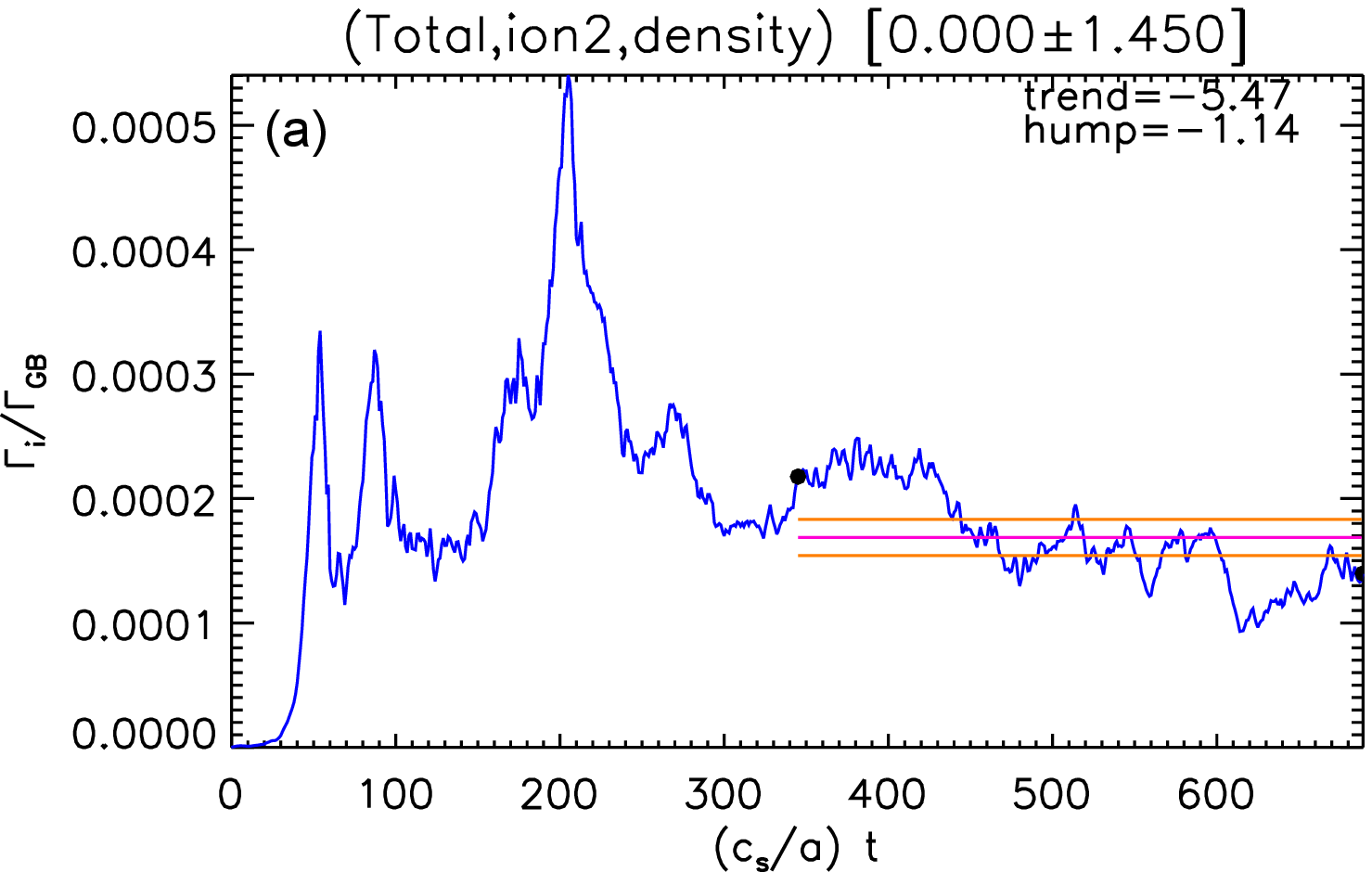}
\end{minipage}%
\begin{minipage}{.5\textwidth}
\includegraphics[height=0.58\textwidth]{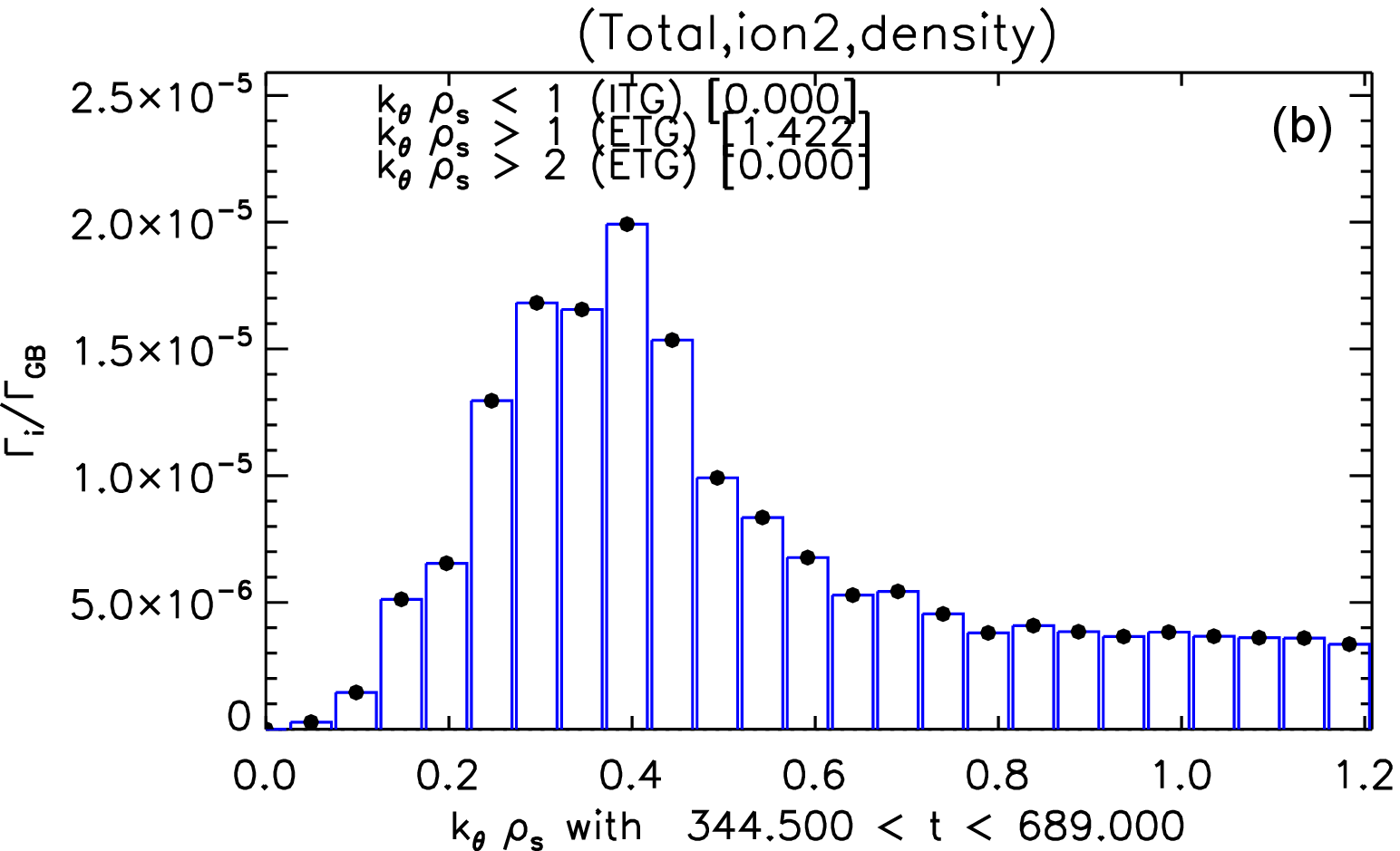}
\end{minipage}
  \caption{$\mathrm{Mo}^{+32}$ fluxes from \gyro~simulations at $r/a=0.56$.
  (a)~Time-evolution of the fluxes. (b)~ Poloidal wave number spectra from the time when the simulations have reached a saturated state.}
\label{fig:NonlinearGYROImpurities}
\end{figure}

\begin{figure}[!ht]
\includegraphics[width=0.49\textwidth]{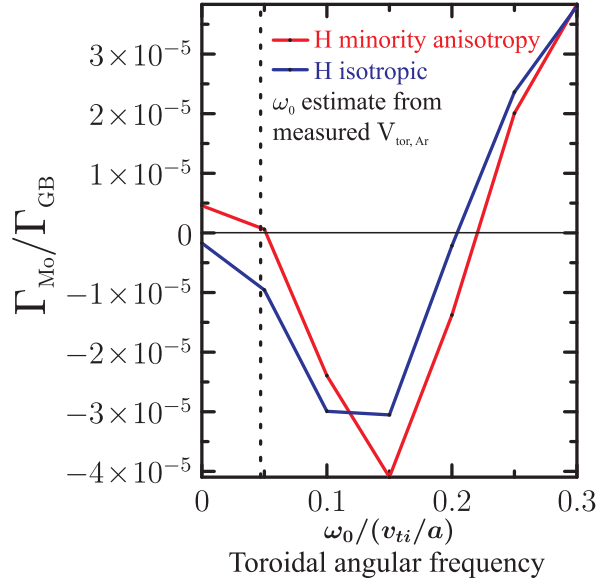}
  \caption{\new{$\mathrm{Mo}^{+32}$ fluxes from \neo~simulations at $r/a=0.56$, as function of toroidal angular rotation frequency $\omega_0$. 
  The experimentally estimated value of $\omega_0$ is marked by a dotted vertical line. 
  The blue line shows results from simulations with isotropic H minority species, while the red line shows results from simulations 
  when the temperature anisotropy from ICRH is taken into account.}}
\label{fig:NEOfluxes}
\end{figure}

To estimate if turbulent impurity fluxes dominate over neoclassical
fluxes, we artificially introduce $\mathrm{Mo}^{+32}$ in contents of
$n_z / n_e = 2 \times 10^{-4}$ at the studied flux surfaces and
perform nonlinear \gyro~simulations. %and \neo~simulations \cite{neo}. 
The
$\mathrm{Mo}^{+32}$ fluxes from the \gyro~simulations at $r/a=0.56$
are shown in Fig.~\ref{fig:NonlinearGYROImpurities}, note that the
largest impurity fluxes are also driven at $k_y \rho_s = 0.40$.  
\new{
We have also performed nonlinear \gyro~simulations with a different ion 
composition consisting of
$\mathrm{B}^{+5}$, $\mathrm{Ar}^{+18}$ and $\mathrm{Mo}^{+32}$ in
contents of $n_{\mathrm{B}} / n_e = 0.02$, $n_{\mathrm{Ar}} / n_e =
10^{-3}$ and $n_{\mathrm{Mo}} / n_e = 2 \times 10^{-4}$, at $r/a=0.56$ and for two different values of $a/L_{nz}$. 
This mix of impurities has similar $Z_{\mathrm{eff}}$ as observed in the experiment, and the combined effects of the dilution result in predicted neutron rate similar to measurements, within uncertainties of the ion temperature. 
The $\mathrm{Mo}^{+32}$ fluxes from these simulations are of the same size as 
the fluxes shown in Fig.~\ref{fig:NonlinearGYROImpurities}. 
Thus we make the conclusion that the magnitude of the turbulent $\mathrm{Mo}^{+32}$ 
fluxes are not sensitive to uncertainties in the ion composition.
} 
\new{
Neoclassical fluxes have been calculated with \neo~\cite{neo} simulations, both including 
and excluding the temperature anisotropy in the H minority from ICRH as well as plasma rotation, 
and varying the value of $a/L_{nz}$. 
Figure~\ref{fig:NEOfluxes} shows the neoclassical $\mathrm{Mo}^{+32}$ fluxes as functions  
of toroidal angular rotation frequency $\omega_0$, where the experimentally estimated value 
is marked by a dotted vertical line. The fluxes are shown for both the case when the 
temperature anisotropy in H minority is included and the case when it is excluded in the simulation. 
%There is no considerable difference between the two cases. 
There is no qualitative difference between the rotation dependence of the neoclassical $\mathrm{Mo}^{+32}$ flux for the two cases. 
The neoclassical fluxes are sensitive to plasma rotation, but found to be very small for the 
experimentally estimated value. Comparing the neoclassical fluxes to the turbulent we find that
$\Gamma_{\mathrm{Mo}}^{\mathrm{GYRO}}/\Gamma_{\mathrm{Mo}}^{\mathrm{NEO}} \gtrsim 10$ 
irrespective of the rotation frequency in the \neo~simulations. 
Furthermore, we note that both the neoclassical pinch velocity and diffusivity (in all \neo~simulations performed) 
are found to be at least an order of magnitude 
smaller than the corresponding turbulent contributions (from \gyro~simulations) to the $\mathrm{Mo}^{+32}$ fluxes. 
{\sc tglf} \cite{tglf} simulations show that if the plasma rotation is assumed to be twice the experimental value, 
turbulent $\mathrm{Mo}^{+32}$ fluxes could be reduced to the same level as the neoclassical.
}
%We
%find that the turbulent fluxes dominate by more than a factor 100 at
%both radial locations, and it should thus not lead to a substantial
%loss in accuracy to neglect the neoclassical fluxes in the model.

Tables~\ref{tab:ra0p38} and \ref{tab:ra0p56} show the ion and electron
energy fluxes ($Q_i$ and $Q_e$, respectively) and the electron
particle flux $\Gamma_e$ normalized to gyro-Bohm units
($Q_{\mathrm{GB}} = n_e c_s T_e \rho_\ast^2$ and $\Gamma_{\mathrm{GB}}
= n_e c_s \rho_\ast^2$ where $\rho_\ast = \rho_s / a$). The turbulent
fluxes calculated by nonlinear \gyro~simulations and the neoclassical
values obtained from \neo~simulations are compared to the result of
power balance calculations representing the experimental value. The
power balance energy fluxes were obtained by \transp~simulations and
they are estimated to have an uncertainty of $\pm 25\%$. Since there
is no core fuelling, the particle sources at these locations are
negligibly small, accordingly, the experimental value of $\Gamma_e$ is
taken to be zero. The neoclassical fluxes are much lower than the
simulated turbulent fluxes, as expected for tokamak mid-radius
parameters.

The turbulent ion energy fluxes at the nominal value of $a/L_{Ti}$ are
significantly higher than the experimental fluxes. However, the
transport is very stiff, i.e. it is sensitive to small changes in
local plasma parameters, especially $a/L_{Ti}$, as it is often
observed in ITG dominated plasmas. Tables~\ref{tab:ra0p38} and
\ref{tab:ra0p56} also show the fluxes at $a/L_{Ti}\pm 10\%$. Without
performing a full sensitivity study to different plasma parameters, as
that is outside the scope of the present paper, from the observed
stiffness of the transport, it is reasonable to believe that the ion
energy fluxes can be matched by reducing $a/L_{Ti}$ and adjusting
other parameters, including impurity composition, within their
experimental uncertainty (noting that $a/L_{Ti}$ has an uncertainty of
roughly $20\%$). The \gyro~electron energy fluxes are comparable to
the experimental values already at nominal $a/L_{Ti}$, however it
seems unlikely that both $Q_i$ and $Q_e$ could be matched to
experiment simultaneously without performing a multi-scale gyrokinetic
analysis.  A stronger electron than ion heat transport channel has
previously been observed in ICRH heated C-Mod experiments
\cite{porkolab,howard}. Also, main ion dilution can significantly
reduce the ion energy fluxes, as found in Ref.~\cite{porkolab}. 
\new{
In the nonlinear \gyro~simulations at $r/a=0.56$ described earlier, 
including the lower-$Z$ (more diluting) impurities $\mathrm{B}^{+5}$ and $\mathrm{Ar}^{+18}$, 
we find that the energy fluxes are significantly reduced. The values are given in 
Tab.~\ref{tab:ra0p56}. $Q_i$ becomes substantially closer to its experimental value, 
whereas $Q_e$ is reduced to lower than the experimental value.
}
%, while in our
%nonlinear simulations, low-$Z$ (more diluting) impurities are not
%included. 
To study the impurity transport using our linear model, it
is not necessary to use the exact parameters that would match the
experimental fluxes, since the impurity peaking factor calculated in
our linear model presented in the next section is only weakly
sensitive to the mode characteristics.

\begin{minipage}{0.9\textwidth}
\begin {table}[H]
 %\caption {Table Title}
 %\label{tab:ra0p38}
 %\begin{center}
  \begin{tabular}{ >{\raggedright\arraybackslash}m{1.5in}  >{\centering\arraybackslash}m{1.5in} >{\centering\arraybackslash}m{1.25in} >{\centering\arraybackslash}m{1.0in} %>{\centering\arraybackslash}m{1.0in} >{\centering\arraybackslash}m{.75in}
  }
   \toprule[1.5pt]
{\textbf{Case}} & $Q_i~\left[Q_{\mathrm{GB}}\right]$ & $Q_e~\left[Q_{\mathrm{GB}}\right]$ &  $\Gamma_e ~\left[\Gamma_{\mathrm{GB}}\right]$\\
   \midrule
   \gyro~$a/L_{Ti}$ -10\%       &   $0.895 \pm 0.098$    & $0.18 \pm 0.02$      & $0.012 \pm 0.001$    \\
  \textbf{GYRO}        &   $\mathbf{1.58} \pm 0.092$  & $\mathbf{0.32} \pm 0.018$      &  $\mathbf{0.017} \pm 0.002$      \\
   \gyro~$a/L_{Ti}$ +10\%       &   $2.03 \pm 0.18$    & $0.40 \pm 0.034$      & $0.024 \pm 0.003$     \\
   \neo       &   $9.2 \times 10^{-3}$   & $4.2 \times 10^{-4}$      & $9.3 \times 10^{-5}$      \\
   Power balance      &    $0.084$    & $0.28$       &  $0$   \\
  \bottomrule[1.25pt]
  \end {tabular}
  \\[0.5pt] %You can adjust how far below the table the text should appear
  %Is just like a caption
  \caption {Comparison of simulated particle- and energy fluxes from \gyro~(turbulent) and \neo~(neoclassical), and experimentally measured fluxes, at $r/a = 0.38$. Also given are \gyro~fluxes when $a/L_{Ti}$ has been artificially increased or decreased by 10\%.
  Fluxes are given in Gyro-Bohm units and $\rho_{\ast} = 5.4 \times 10^{-3} $. Note that positive (negative) flux is outward (inward).}
 %\end{center}
 \label{tab:ra0p38}
\end {table}
\vfill
\end{minipage}

\begin{minipage}{0.9\textwidth}
\begin {table}[H]
 %\caption {Table Title}
 %\label{tab:ra0p56}
 %\begin{center}
  \begin{tabular}{ >{\raggedright\arraybackslash}m{1.5in}  >{\centering\arraybackslash}m{1.5in} >{\centering\arraybackslash}m{1.25in} >{\centering\arraybackslash}m{1.0in} %>{\centering\arraybackslash}m{1.0in} >{\centering\arraybackslash}m{.75in}
  }
   \toprule[1.5pt]
{\textbf{Case}} & $Q_i~\left[Q_{\mathrm{GB}}\right]$   & $Q_e~\left[Q_{\mathrm{GB}}\right]$ & $\Gamma_e ~\left[\Gamma_{\mathrm{GB}}\right]$ \\
   \midrule
  \gyro~$a/L_{Ti}$ -10\%       &   $1.20 \pm 0.032$     & $0.81 \pm 0.016$      &  $-0.22 \pm 0.004$   \\
   \textbf{GYRO}        &    $\mathbf{1.96} \pm 0.07$    & $\mathbf{1.23} \pm 0.04$      &    $\mathbf{-0.29} \pm 0.007$ \\
   \gyro~$a/L_{Ti}$ +10\%       &   $3.56 \pm 0.12$  & $2.18 \pm 0.088$      &   $-0.38 \pm 0.014$      \\
   \new{\gyro~$\mathrm{B}$/$\mathrm{Ar}$/$\mathrm{Mo}$}      &   \new{$ 0.64 \pm 0.046$}  & \new{$ 0.38 \pm 0.031 $}      &   \new{$ -0.13 \pm 0.008$}      \\
   \neo       &  $0.013$    & $1.1 \times 10^{-3}$      &   $1.8 \times 10^{-4}$    \\
   Power balance    &  $0.38$   &  $0.88$    &   $0$     \\
  \bottomrule[1.25pt]
  \end {tabular}
  \\[0.5pt] %You can adjust how far below the table the text should appear
  %Is just like a caption
  \caption {Comparison of simulated particle- and energy fluxes from \gyro~(turbulent) and \neo~(neoclassical), and experimentally measured fluxes, at $r/a = 0.56$. Also given are \gyro~fluxes when $a/L_{Ti}$ has been artificially increased or decreased by 10\%, 
  \new{
  and when the ion composition has been modified to include $\mathrm{B}^{+5}$, $\mathrm{Ar}^{+18}$ and $\mathrm{Mo}^{+32}$ in
contents of $n_{\mathrm{B}} / n_e = 0.02$, $n_{\mathrm{Ar}} / n_e =
10^{-3}$ and $n_{\mathrm{Mo}} / n_e = 2 \times 10^{-4}$.
  }
  Fluxes are given in Gyro-Bohm units and $\rho_{\ast} = 4.4 \times 10^{-3} $. Note that positive (negative) flux is outward (inward).}
 %\end{center}
 \label{tab:ra0p56}
\end {table}
\vfill
\end{minipage}

%%%%%%%%%%%%%%%%%%%%%%%%%%%%%%%%%%%%%%%%%%%%%%%%%%%%%%%%%%%%%%%%%%%%%%%%%%%%%%%%%%%%%%%%%%%%%%%%%%%%%%%%%%%%%%%%%%%%%%%%%%%%%%%%%%%%%%%%%%%%%%
%%%%%%%%%%%%%%%%%%%%%%%%%%%%%%%%%%%%%%%%%%%%%%%%%%%%%%%%%%%%%%%%%%%%%%%%%%%%%%%%%%%%%%%%%%%%%%%%%%%%%%%%%%%%%%%%%%%%%%%%%%%%%%%%%%%%%%%%%%%%%%

\section{Impurity density peaking}\label{sec:impurity}

A semi-analytical linear gyrokinetic model for the zero flux density
gradient of impurities present in trace quantities was introduced in
Ref.~\cite{fulop} and later extended to include parallel streaming
effects in the high-$Z$ limit in Ref.~\cite{varenna}, where the effect of
a poloidally varying, non-fluctuating electrostatic potential $\phi_E$
is included with $e_z \phi_E / T_z \sim \ord \left(1\right)$. From
Fig.~\ref{fig:VaryingPotential} we note that this ordering is valid
for the studied ICRH discharge.  In Ref.~\cite{albertTEM} the model
was applied to TEMs, and demonstrated the ability to reproduce trends
of nonlinear \gyro~simulations in the poloidally symmetric case. An
important result was that the strength of the asymmetry $\mathcal{K}$
and the magnetic shear $s$ are key parameters in determining the size
of the effect, since they both appear as explicit factors in the
$\Ev\times\Bv$ drift which arises due to the poloidal variation in the
potential.  References \cite{varenna,albertTEM} assumed a sinusoidal
variation in the non-fluctuating potential of the form in
Eq.~(\ref{eq:Poloidal_potential}), but as discussed in
Sec.~\ref{sec:experimental} we need to allow for a more general form
in the present analysis.  Furthermore, similarly to Ref.~\cite{varenna} we
assume a low-beta plasma and large-aspect-ratio, but in the present
treatment we will allow for shaping effects in form of plasma
elongation, $\kappa$. 
In Ref.~\cite{skyman2014} the impact of shaping effects on the impurity peaking factor 
is investigated, and it is found that considering a realistic geometry can significantly 
modify the high-$Z$ peaking factor as compared to using a simplified circular geometry. 
Furthermore it is argued that the main effect can be attributed to elongation.
 We note that employing Miller type
parametrization an elliptical flux surface with elongation $\kappa$ is
given above Eq.~(\ref{mj}), where $\theta$ is not the geometrical
poloidal angle $\Theta$, but an angle-like parameter which also varies
as $\theta: 0 \rightarrow 2\pi$, and is related to $\Theta$ by $ \tan
\Theta = \kappa \tan \theta
%\label{eq:RelationTTheta}
%\end{equation}
$. 

The linearized gyrokinetic equation for the non-adiabatic perturbed
impurity distribution $g_z$ is given by
\begin{equation}
  \left.\frac{v_\parallel}{q R} \frac{\partial {g}_z}{\partial
    \theta}\right|_{\mathcal{E},\mu}
    -i(\omega- \omega_{Dz}-\omega_E) {g}_z -
  C[g_z] =-i\frac{Z e f_{z0}}{T_z}\left(\omega-\omega_{\ast
    z}^T \right)\phi J_0(z_z).
\label{gke}
\end{equation}
Here $f_{z0}=n_{z0}(m_z/2\pi T_z)^{3/2}\exp(-\energy/T_z)$ is the
equilibrium Maxwellian distribution function, $\energy = m_z v^2/2 + Z
e \phi_E$ is the total unperturbed energy, $\mu = m_z \vpe^2/\left(2
B\right)$ is the magnetic moment, $n_z(\mathbf{r})=n_{z0}\exp[-Z
  e\phi_E(\mathbf{r})/T_z]$ is the poloidally varying impurity density
where $n_{z0}$ is a flux function, and $\phi$ is the perturbed
potential.  $J_0$ is the Bessel function of the first kind,
$z_z=k_\perp v_{\perp}/\Omega_{z}$, and $k_\perp = \left(1 + s^2
\theta^2\right)^{1/2} k_y$.  $ \omega_{Dz}=
% -2 k_\theta /m_z(\energy-Z e\phi_E-\mu B/2) \mathcal
% D(\theta)/\omega_{cz} R
-2 k_y T_z (x_{\perp}^2/2+x_{\parallel}^2) \mathcal
D\left(\theta\right)/\left(m_z\Omega_{z} R \kappa\right) $ is the
magnetic drift frequency, where $\mathcal D\left(\theta\right)=
\cos{\theta}+ s \theta \sin{\theta}$ and $x=v/v_{Tz}$ represents
velocity normalized to the thermal speed $v_{Tz}=(2T_z/m_z)^{1/2}$,
whereas $\omega_{\ast z}=-k_y T_z/Z eB \kappa L_{nz}$ is the diamagnetic
frequency and $\omega_{\ast z}^T = \omega_{\ast z}\left[1 %- L_{nz} Ze \partial \phi_E/\partial r/T_z
  +\left(x^2-3/2\right)L_{nz}/L_{Tz}\right]$.  $\omega_E$ is the
$\Ev\times \Bv$ drift frequency of the particles in the
non-fluctuating electrostatic field given by
\begin{equation}
\omega_{E}=-\frac{k_y}{B} \frac{s \theta}{r \kappa} \frac{\partial
  \phi_E}{\partial \theta},
\label{ome}
\end{equation}
if radial variation of $\phi_E$ is neglected, the term was derived in Appendix A
of Ref.~\cite{albertICRH}.  $C[\cdot]$ is the collision operator.

The impurity peaking factor is found by solving Eq.~(\ref{gke}) and
requiring that the linear impurity flux $\Gamma_z$ should vanish
\begin{equation}
  0 = \left\langle\Gamma_z\right\rangle \equiv \left\langle
  \mathrm{Im} \left[ -\frac{k_y}{B}\hat{n}_z\phi^\ast \right]
  \right\rangle = \left\langle \mathrm{Im} \left[-\frac{k_y}{B}\int
    d^3v J_0\left(z_z\right) g_z \phi^\ast \right] \right\rangle
  .\label{ge}
\end{equation}

A perturbative solution to Eq.~(\ref{gke}) in the small parameter
$Z^{-1/2} \ll 1$ was presented in Ref.~\cite{varenna}, keeping terms up to
$\ord(Z^{-1})$ in the expansion of ${g}_z$.  The solution orders
$\omega_{Dz}/\omega$, $\omega_{\ast z}^T/\omega$, $\omega_E/\omega$
and $J_0(z_z)-1\approx -z_z^2/4$ as $\sim\!1/Z$ small (assuming
$m_z/m_i\sim Z$), and models collisions by the full linearized
impurity-impurity collision operator $C_{zz}^{(l)}$. Furthermore it is
also assumed that $\phi$ and $\omega$ are known from the solution of
the linear gyrokinetic-Maxwell system (obtained from \gyro) and that
they are unaffected by the presence of trace impurities, and the
non-fluctuating potential.  We refer to Ref.~\cite{varenna} for more
details. A semi-analytical expression for the impurity peaking factor
is given in Eq.~(8) of Ref.~\cite{varenna}.  In the present analysis
we will adopt the same perturbative solution, but keep the final
solution for the peaking factor numerical to account for a more
complicated non-fluctuating potential and $\ord \left(\epsilon\right)$
corrections. Since it is shown in Ref.~\cite{varenna} that collisions do
not affect impurity peaking to order $1/Z$, we will neglect them here
for simplicity.  Writing $g_z = g_0 + g_1 + g_2 + \ord
\left(1/Z^{3/2}\right)$, the 0${}^{\mathrm{th}}$ order solution is
given by $g_0 = Z e \phi f_{z0} / T_z$, the 1${}^{\mathrm{st}}$ order
solution by $g_1 = -i Z e f_{z0} \vpa %\p_{\theta} \left(\phi\right)
\left(\p \phi / \p \theta \right)
 / \left(T_z \omega q R\right)$, and the 2${}^{\mathrm{nd}}$ order
solution by $g_2 = \left(\omega_{Dz} + \omega_{E} \right) g_0 / \omega
+ \vpa %\p_{\theta} \left(g_1\right) 
\left(\p g_1 / \p \theta\right)
/ \left(i \omega q R\right) - Z e
\phi f_{z0} \left(\omega z_z^2/4 + \omega_{\ast z}^T\right) /
\left(\omega T_z\right)$.  Since $g_0$ is an adiabatic response and
$g_1$ disappears when integrating over velocity space, only $g_2$
contributes to the impurity fluxes.  Furthermore, since the impurity
density varies over the flux surface, what we calculate is an
effective impurity peaking factor $a / L_{nz}^0 = a / L_{nz}^\ast$
when $\left\langle\Gamma_z\right\rangle = 0$, where $a / L_{nz}^\ast =
\langle a / L_{nz} \rangle_\phi$ is a weighted flux surface average
with $\langle\dots\rangle_\phi=\langle\dots \mN
|\phi|^2\rangle/\langle\mN |\phi|^2\rangle$ and $\mathcal
N(\theta)=\exp{[- e_z \phi_E / T_z]}$.

Regarding the effects of a constant elongation, we note that
$\kappa>1$ reduces the magnetic, the non-fluctuating $\Ev\times\Bv$
and the diamagnetic drift terms by a factor $1/\kappa$, while it does
not %effect 
\new{affect} 
the parallel streaming term, as shown in Appendix A. In
terms of the impurity peaking factor, the elongation increases the
parallel streaming term by a factor $\kappa$, but it leaves the drift
contributions unchanged. Thus, in ITG dominated plasmas where parallel
streaming acts as to increase the peaking factor, the peaking factor
is expected to be higher than in a similar circular cross section
plasma, especially in the core where $q^2$ is not too large.

We consider a single representative linear mode, which is the mode at
$k_y \rho_s = 0.40$ driving the largest fluxes nonlinearly (see
Fig.~\ref{fig:NonlinearGYROImpurities}).  Note that since we are using
a linear model, only the most unstable mode is considered and any
sub-dominant modes are neglected. At $r/a = 0.56$ there exists a
sub-dominant TEM with very low growth rate (see
Fig.~\ref{fig:Eigenvalues2D}). From a potential frequency response
analysis of the nonlinear simulations we found that fluctuations with
frequencies in the electron diamagnetic direction have negligibly
small amplitudes, thus we do not expect that neglecting the transport
driven by the sub-dominant TEM affects the results significantly.

\begin{figure}[!ht] \scalebox{0.95}{\includegraphics[width=0.49\textwidth]{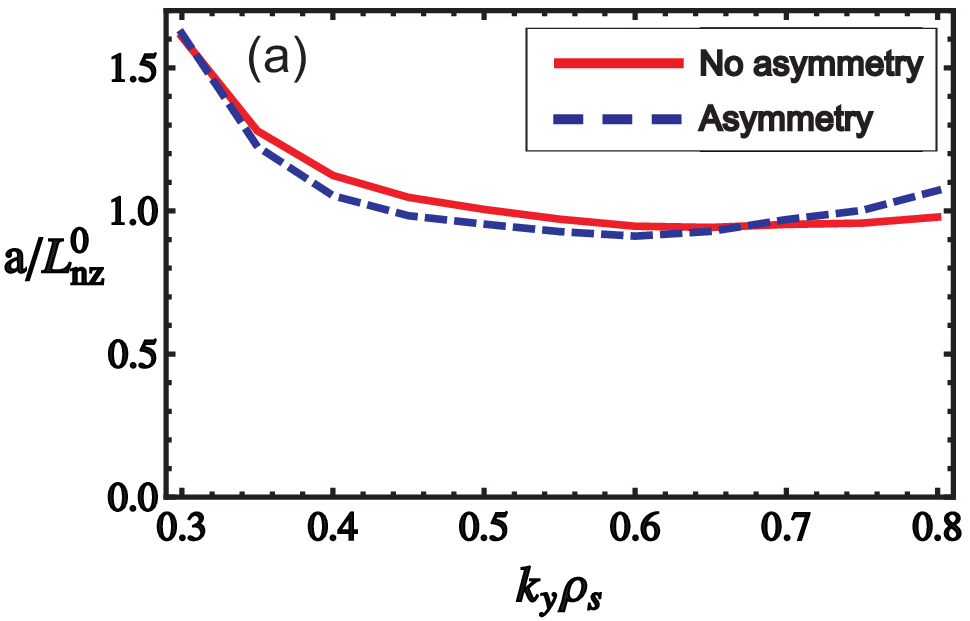}
\includegraphics[width=0.49\textwidth]{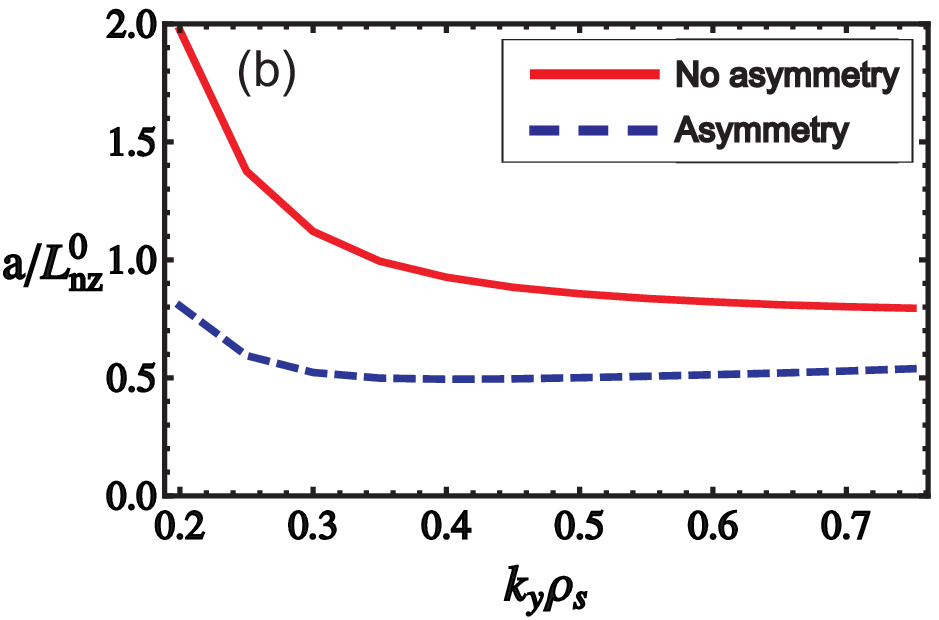}} 
  \caption{$\mathrm{Mo}^{+32}$ peaking factor at $r/a=0.38$ (a) and at $r/a=0.56$ (b) as functions of $k_y \rho_s$.
  Red solid line is the peaking factor when the poloidally varying non-fluctuating potential is excluded from the model, while blue dashed line is the peaking factor when the potential is included.
}
\label{fig:ModelPeakingFactors}
\end{figure}
Figure~\ref{fig:ModelPeakingFactors} shows the model
$\mathrm{Mo}^{+32}$ peaking factor at $r/a=0.38$ and $r/a=0.56$ as
function of binormal wave number, when the poloidally varying
non-fluctuating potential is excluded and included in the model. For
$k_y \rho_s = 0.40$, the mode driving the largest fluxes, the peaking
factor is practically unaffected by the inclusion of the potential at
$r/a=0.38$, while at $r/a=0.56$ it decreases from $\new{a/L_{nz}^0} = 0.93$ to
$\new{a/L_{nz}^0} = 0.49$.  One of the reasons for the weak effect at
$r/a=0.38$ is a moderate value of the magnetic shear $s = 0.33$, which
makes the $\mathbf{E}_{\theta} \times \mathbf{B}_{\varphi}$-drift term
relatively small.  At $r/a=0.56$ the magnetic shear is $s = 0.78$ and
the $\mathbf{E}_{\theta} \times \mathbf{B}_{\varphi}$-drift is
stronger. However, the approximate proportionality of the
$\mathbf{E}_{\theta} \times \mathbf{B}_{\varphi}$-drift to magnetic
shear in itself cannot explain the large relative increase in the
impact of poloidal asymmetries. Figure~\ref{fig:ModelPeakingFactors}
is calculated retaining $\ord(\epsilon)$ corrections to the peaking
factor. We note, that the reduction of the peaking factor due to
poloidal asymmetry effects is found to be much smaller when a
simplified model neglecting $\ord(\epsilon)$ corrections is used for
the calculation. This suggests that these corrections may be important
for some set of parameters, as it is the case here.

\begin{figure}[!ht]
\includegraphics[width=0.49\textwidth]{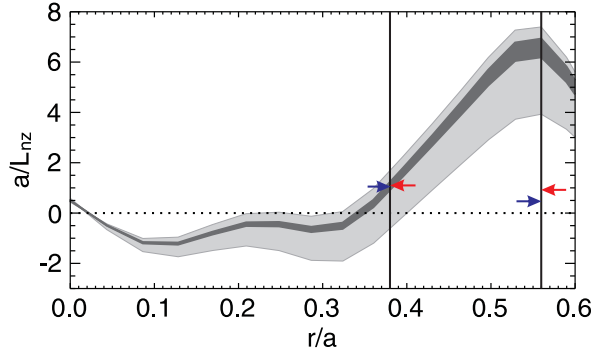}
  \caption{ Radial variation of the impurity peaking factor. Light and
    dark %green 
    \new{gray} 
    shaded areas represent the experimental values
    calculated using a less and a more restricting set of assumptions
    on the uncertainties. Vertical bars mark the two studied
    radii. Arrows represent the theoretical estimates for the peaking
    factor at these radial locations; from the left (blue arrows):
    with asymmetries, from the right (red arrows): without
    asymmetries.}
\label{fig:ExplPeakingFactors}
\end{figure}

Figure~\ref{fig:ExplPeakingFactors} shows the comparison of the
peaking factors as calculated above to the experimental range. The
codes GENTRAN \cite{gentran} and STRAHL \cite{strahl} were used to
constrain the impurity profile from the measured emission data
\cite{mattAPS}. GENTRAN solves for the steady state charge state
density profiles given assumed input impurity diffusion and convection
profiles. We assume a flat diffusion profile, and the convection
profile is found using a least-squares minimization routine that best
matches the time-averaged neon-like molybdenum emissivity profile for
$0.10 < r/a < 0.53$.  Assuming a finite range for diffusivities, 0.5
to 1.8~$\mathrm{m}^2/\mathrm{s}$, and considering the uncertainties in
the contribution due to charge exchange recombination provides a range
for the experimental peaking factors can be calculated that are
consistent with experiment. 
 Figure~\ref{fig:ExplPeakingFactors} shows two uncertainty ranges for
 the impurity peaking factor. The light gray area is based purely on
 GENTRAN simulations and use less restrictive assumptions, allowing
 larger uncertainties. At the same time, the diffusion and convection
 profiles calculated by GENTRAN can be used as inputs for STRAHL
 simulations.  STRAHL computes the time-evolving charge state density
 profiles, in particular, it can predict the exponential decay times
 of the impurity density during the laser blow off experiment.  The
 dark-gray region in Fig.~\ref{fig:ExplPeakingFactors} reflects the
 molybdenum density profiles for which the decay times predicted
 by STRAHL are within the experimental uncertainty.  We note that,
 although it would be ideal to use STRAHL alone to simulate the full
 time history and constrain the diffusion and convection profiles,
 uncertainties in the atomic physics data for molybdenum currently
 prevent using this proven work flow.  The theoretically predicted
 linear peaking factors (corresponding to $k_y \rho_s = 0.40$) at the
 two studied radial locations are shown with arrows. The arrows from
 the left (blue) include the poloidal asymmetry effects, while the
 arrows from the right are calculated assuming no ICRH driven poloidal
 asymmetries (these correspond to the blue and red curves of
 Fig.~\ref{fig:ModelPeakingFactors}, respectively).

 At $r/a=0.38$, the calculated peaking factors -- which, in this case,
 are almost the same with and without asymmetries -- fall within the
 more restrictive experimental range. However, the ones calculated for
 $r/a=0.56$ strongly underestimate the experimental impurity peaking
 factor, even when the less restrictive range is considered. Note
 that, although we neglected rotation effects in the calculation, this
 level of discrepancy likely cannot be explained by rotation effects.
 Using the model presented in Ref.~\cite{angioni2012}, drift
 contributions due to rotation on the impurity peaking are estimated
 to be an order of magnitude smaller then the peaking factors
 calculated here.  As seen in Fig.~\ref{fig:Asym}, at this radial
 location the poloidal variation of the impurity density is still not
 dominated by rotation effects, although these become comparable to
 the ICRH-induced effects.  Accordingly the ``no asymmetry'' result
 (red arrow from right in Fig.~\ref{fig:ExplPeakingFactors}) may be
 considered as an estimate for the peaking factor in this case.
% \todo{
% Comment from Reinke:
% Your model only handles poloidal electric field effects while the LFS accumulation is due to centrifugal force (which has a poloidal electric field component due to different ion and electron mass).  You can't exclude rotation effects because the asymmetry is smaller than the ICRH-driven one.
% }
 As long as we assume that the linear peaking
 factor is a good approximation to the nonlinear one, the strong
 discrepancy observed makes it clear that in the outer radial location
 the impurity peaking is determined by some physical mechanism 
 %other than turbulent impurity transport. 
 not included in our model.
% \todo{
% Matt wants to replace ''other
% than turbulent impurity transport''
% with
% ''not included in our model.  One possibility could be centrifugal-drift driven turbulent transport as discussed in [C. Angioni, et al. Phys. Plasmas 19, 122311 (2012)]''
% }
 To obtain such a large peaking
 factor as seen for $r/a=0.56$, there should be a drift frequency 
% \todo{
% Matt: It's worth checking Clemente's analytical model which should estimate the level of the centrifugal drift on the transport.
% }
 in
 the gyrokinetic equation affecting the impurity dynamics that is
 several times larger than the diamagnetic drift frequency.

The discrepancy between the experiment and the simulation at the outer
radius %is likely 
might be 
due to the increasing importance of atomic physics
processes. In Ref.~\cite{mattAPS} the neutral fraction (which is not a
measured quantity) is shown to have a significant impact on the radial
distribution of molybdenum with $Z>30$, especially
outside $r/a=0.5$. 
%\todo{
%An increased neutral fraction, increasing the total recombination rate, is shown to improve agreement on C-Mod, an approach which is qualitatively similar to the need to increase the recombination rates for tungsten on ASDEX-U [T Pütterich et al 2008 Plasma Phys. Control. Fusion 50]
%}
%\todo{
%Matt wants to remove the following two sentences
%}
%Furthermore, the atomic physics calculation has
%only a limited model for the radial transport. In our case a constant
%diffusivity is assumed, which is shown to have difficulties in
%reproducing the line average time history of certain charge
%states~\cite{mattAPS}. 
On the other hand, 
the gyrokinetic modeling of
the impurity transport considers only a single charge state of the
impurity and neglects volumetric sources due to ionization and
recombination. Consequently it is, although important to point out,
not completely unexpected that we do not find agreement between
experiment and modeling at the outer core.
This does illustrate the role of uncertainties in atomic physics data 
in complicating efforts at validating gyrokinetic models of high-$Z$ impurity transport.

\subsection*{Sensitivity analysis of impurity peaking}

\begin{figure}[!ht] 
%\scalebox{0.95}{\includegraphics[width=0.49\textwidth]{figure11a.eps}
%\includegraphics[width=0.49\textwidth]{figure11b.eps}}
\scalebox{0.95}{\includegraphics[width=0.49\textwidth]{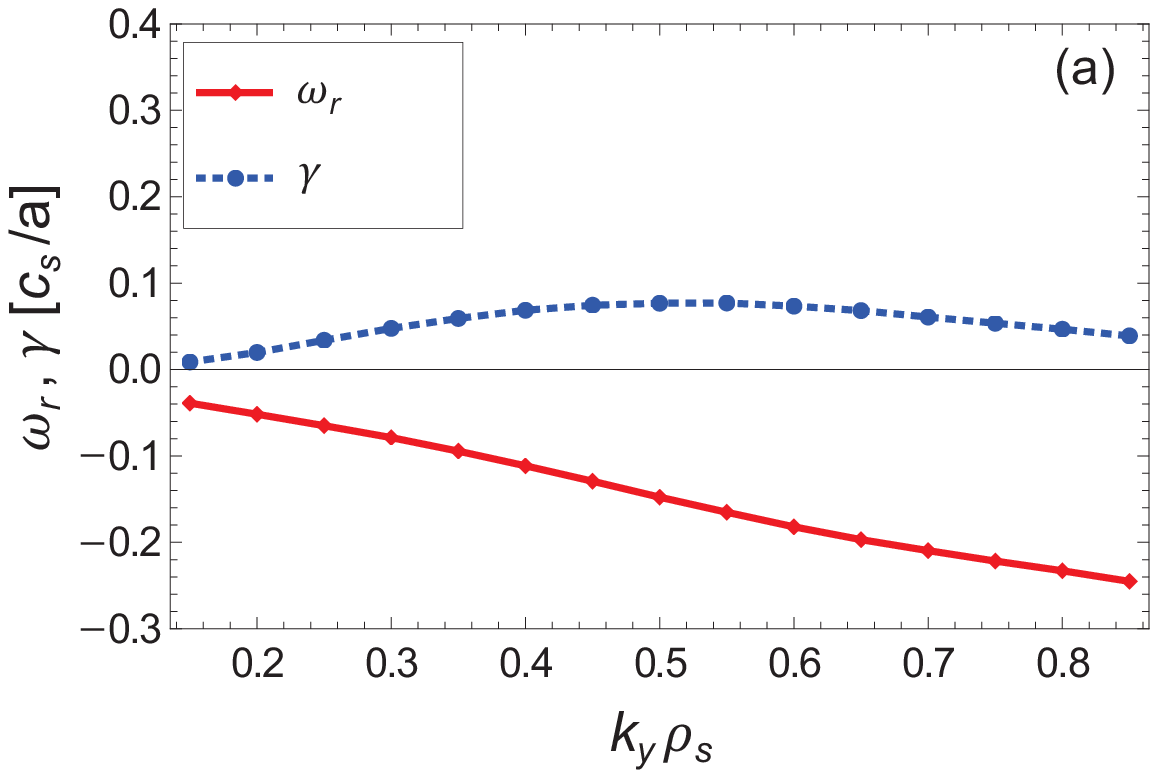}
\includegraphics[width=0.49\textwidth]{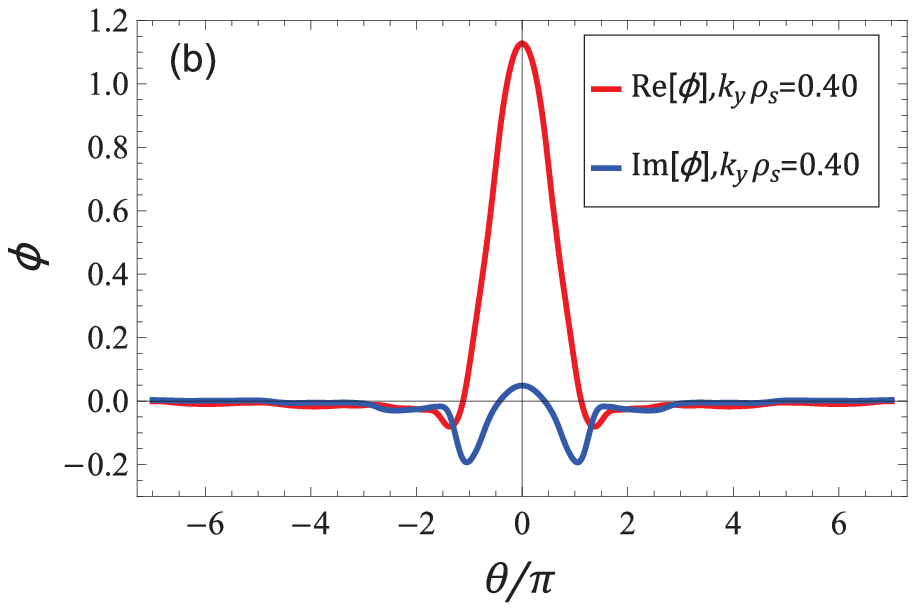}}
  \scalebox{0.95}{%\includegraphics[width=0.49\textwidth]{figs/Alcator_2012_aLnz0_radius0p56_OutsideICRH_SEVERAL_IMP_SPECIES_EquilibriumPotential_new.eps}
    \includegraphics[width=0.49\textwidth]{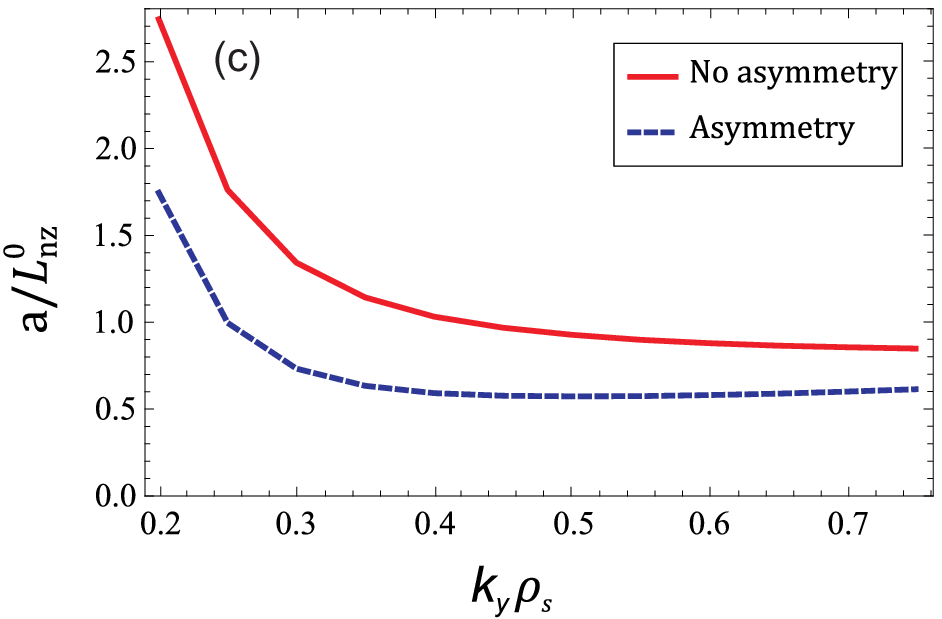}
    \includegraphics[width=0.49\textwidth]{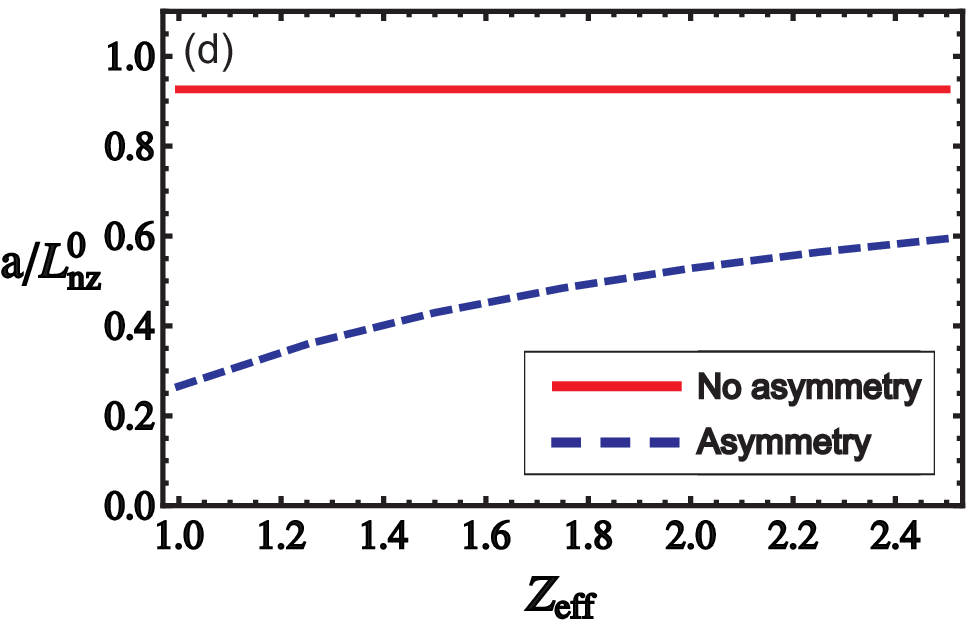}
    }
\caption{(a)~Real mode frequency $\omega_r$ (red solid line with
  diamonds) and linear growth rate $\gamma$ (blue dashed line with
  dots) as functions of $k_y \rho_s$ at $r/a = 0.56$, for a plasma
  with $\mathrm{B}^{+5}$, $\mathrm{Ar}^{+18}$ and $\mathrm{Mo}^{+32}$
  introduced in trace quantities and $Z_{\mathrm{eff}} = 1.90$.
  (b)~Real (red line) and imaginary (blue line) part of the perturbed
  potential as functions of extended poloidal angle at $r/a = 0.56$
  and $k_y \rho_s = 0.40$ for the corresponding plasma.
%  (c)~Non-fluctuating potential as function of poloidal angle at
%  $r/a=0.56$ for the corresponding plasma. 
   (c)~$\mathrm{Mo}^{+32}$
  peaking factor at $r/a=0.56$ as function of $k_y \rho_s$ for the
  corresponding plasma.  Red solid line is the peaking factor when the
  poloidally varying non-fluctuating potential is excluded from the
  model, while blue dashed line is the peaking factor when the
  potential is included.  
  (d)~$\mathrm{Mo}^{+32}$ peaking factor at $r/a=0.56$ and $k_y
  \rho_s = 0.40$ as function of $Z_{\mathrm{eff}}$. 
%   Red solid line is
%  the peaking factor when the poloidally varying non-fluctuating
%  potential is excluded from the model, while blue dashed line is the
%  peaking factor when the potential is included.
  }
\label{fig:ModelPeakingFactorsWithZeff}
\end{figure}

In this subsection we investigate the sensitivity of the impurity
peaking factor to the plasma effective charge, by introducing
$\mathrm{B}^{+5}$, $\mathrm{Ar}^{+18}$ and $\mathrm{Mo}^{+32}$ in
contents of $n_{\mathrm{B}} / n_e = 0.02$, $n_{\mathrm{Ar}} / n_e =
10^{-3}$ and $n_{\mathrm{Mo}} / n_e = 2 \times 10^{-4}$, resulting in
a plasma effective charge of $Z_{\mathrm{eff}} = 1.90$.  Including the
above mentioned impurities, Eq.~(\ref{eq:VaryingPotential}) for the
non-fluctuating potential becomes
\begin{equation}
  \frac{Z e \phi_E}{T_z} = \frac{Z \tilde{n}_H / n_{e 0}}{{T_z}/{T_e} + ({T_z}/{T_i}) ({n_{i 0}}/{n_{e 0}}) +  {n_{z 0}Z^2}/{n_{e 0}} +
  {n_{\mathrm{B} 0} Z_{\mathrm{B}}^2 T_z}/\left(n_{e 0} T_{\mathrm{B}}\right) + {n_{\mathrm{Ar} 0} Z_{\mathrm{Ar}}^2 T_z}/\left(n_{e 0} T_{\mathrm{Ar}}\right)}.
\label{eq:VaryingPotentialZeff}
\end{equation}
%\red{(why there is no Mo in this formula ?)} 
The inclusion of the $\mathrm{B}^{+5}$ and $\mathrm{Ar}^{+18}$
impurities lead to a reduction of the non-fluctuating potential, and
accordingly also the peaking factor is slightly increased in the poloidally asymmetric case, as shown in
Fig.~\ref{fig:ModelPeakingFactorsWithZeff}(c) for the radial location
$r/a=0.56$.  Interestingly, the TEM which was dominant for higher wave
number in Fig.~\ref{fig:EigenvaluesPotential} at $r/a = 0.56$
disappears if the above mentioned impurities are included, as illustrated in Fig.~\ref{fig:ModelPeakingFactorsWithZeff}(a). 

%\begin{figure}[!ht]
%  \scalebox{0.95}{\includegraphics[width=0.49\textwidth]{figs/Alcator_2012_aLnz0_radius0p56_ZeffScan_without_withAsymmetry_withElongation_NumericalModel.eps}}
%\caption{$\mathrm{Mo}^{+32}$ peaking factor at $r/a=0.56$ and $k_y
%  \rho_s = 0.40$ as function of $Z_{\mathrm{eff}}$.  Red solid line is
%  the peaking factor when the poloidally varying non-fluctuating
%  potential is excluded from the model, while blue dashed line is the
%  peaking factor when the potential is included.  }
%\label{fig:ModelPeakingFactorZeff}
%\end{figure}

Figure~\ref{fig:ModelPeakingFactorsWithZeff}(d) %\ref{fig:ModelPeakingFactorZeff}
 shows how the
$\mathrm{Mo}^{+32}$ peaking factor varies with $Z_{\mathrm{eff}}$,
under the assumption that the turbulence is unaffected by the change
in impurity content 
which is expected as long as all impurity species are in 
trace quantities.
The figure illustrates that the peaking factor is
sensitive to the effective plasma charge, which is also evident from
Eq.~(\ref{eq:VaryingPotential}).  To be able to predict the magnitude
of the poloidal asymmetry in the non-fluctuating potential a good
estimate of $Z_{\mathrm{eff}}$ is required. However, neither the
inclusion of lower $Z$ impurities or changing the effective charge
produces strong enough deviation in the peaking factor to explain the
experimentally observed very high value.

%%%%%%%%%%%%%%%%%%%%%%%%%%%%%%%%%%%%%%%%%%%%%%%%%%%%%%%%%%%%%%%%%%%%%%%%%%%%%%%%%%%%%%%%%%%%%%%%%%%%%%%%%%%%%%%%%%%%%%%%%%%%%%%%%%%%%%%%%%%%%%
%%%%%%%%%%%%%%%%%%%%%%%%%%%%%%%%%%%%%%%%%%%%%%%%%%%%%%%%%%%%%%%%%%%%%%%%%%%%%%%%%%%%%%%%%%%%%%%%%%%%%%%%%%%%%%%%%%%%%%%%%%%%%%%%%%%%%%%%%%%%%%

\section{Discussion and conclusions}
\label{sec:conclusions}

We have performed a gyrokinetic study of the turbulent transport of
highly charged molybdenum impurity in an off-axis ICRH heated discharge on
Alcator C-Mod. The discharge is part of an experiment series to study
the effect of ICRH induced asymmetries on high-$Z$ impurity transport,
where multiple radiation imaging diagnostics follow the
spatio-temporal dynamics of molybdenum introduced using laser blow-off.

We find that for inner core radii ($r/a<0.5$) there is a significant
ICRH-induced in-out asymmetry, while further out this asymmetry is not
as pronounced 
or driven by centrifugal effects. 
This trend is partly due to the ICRH resonance location
determining the poloidally varying non-fluctuating potential and
partly due to rotation effects, even though this discharge has a low
rotation speed on the diamagnetic level.  We perform eigenvalue solver
gyrokinetic simulations with {\gyro}, resolved radially and in
$k_y\rho_s$, which show unstable microinstabilities for $r/a>0.3$. In
the inner core ($r/a<0.55$) the turbulence is ITG dominated, while in
the outer core an ITG and a higher wave number TEM coexist.

Nonlinear gyrokinetic simulations at the nominal values of plasma
parameters significantly overestimate the ion heat transport at the
studied radial locations, however the transport is found to be rather
stiff. 
\new{
The ion heat transport in the simulations is brought significantly closer to 
the experimental value by modifying the ion composition to include 
lower-$Z$, more diluting, impurities. 
}
The electron energy transport can be matched when the ion
temperature gradient is changed within a $10\%$.

At the inner radius studied, $r/a=0.38$, the impurity peaking factor
calculated using a linear gyrokinetic model matches the experiment
within uncertainties. However the effect of ICRH-induced asymmetries
is negligibly small due to the relatively low magnetic shear. The
difference between the calculated peaking factors with and without
poloidal asymmetries are too close to be experimentally
distinguishable. At the outer radius, $r/a=0.56$, the magnetic shear
is higher and the effect of ICRH is still strong enough to induce a
stronger reduction in the impurity peaking factor. However,
independently on whether poloidal asymmetries are considered or not,
and considering a reasonable sensitivity to ion composition and
gradients, the range of calculated impurity peaking factors
($\new{a/L_{nz}^0}\approx 0.2-1.7$) strongly underestimates the experimentally
observed very high impurity peaking factor ($\new{a/L_{nz}^0}\approx 6$). This
suggests that there is some mechanism in the outer core strongly
affecting impurity transport, %other than turbulent transport. 
 not
presently included in our model.  
\new{Pure neoclassical transport could generate such a large peaking. 
However, 
} 
%Comparing 
comparing 
nonlinear gyrokinetic
simulations and neoclassical simulations we show that neoclassical
transport is too small to be a good candidate. The effects of the weak
rotation in the experiment, which we estimated, are not expected to
qualitatively change the picture either. On the other hand, previous
studies showing a sensitivity of the radial profiles of high
ionization states of molybdenum to neutral fraction indicate that
atomic physics processes
%are playing an important role in the outer core.
may be inducing a systematic error in interpreting experimental results at mid-radius.

%\todo{Matt wants to remove the following paragraph}
%The atomic physics codes currently available have only a limited
%capability to model radial transport. On the other hand, the
%gyrokinetic modeling presented here does not account for any atomic
%physical processes. Our finding implies that there is a need for a
%more integrated approach for a quantitative study of impurity
%transport, which would combine atomic physics and turbulent transport
%modeling.

Although the presented study of an Alcator C-Mod experiment seems to
point towards a conclusion that the ICRH-driven poloidal asymmetry may
not be a tool to control impurity transport, we note that the peaking
factors we find strongly depend on magnetic shear, and are affected by
the radial region where atomic physics becomes important. In larger
devices with hotter plasmas, such as JET or ITER, there might be
bigger room to harvest the favorable effects of ICRH-driven
asymmetries on turbulent transport. In fact, ICRH is routinely applied
on JET to avoid an uncontrolled impurity accumulation event in the
core, and it is still not clear what physics effect plays the dominant
role in that case. Furthermore, in the deep core, where turbulent
transport seems to be absent, the effect of these asymmetries on
neoclassical transport -- which is an area yet to be explored in
detail -- may also be beneficial.

\renewcommand{\theequation}{A\arabic{equation}}
\setcounter{equation}{0}
\section*{Appendix A: Drift frequencies in a large aspect ratio, elongated equilibrium}
\label{sec:appendixA}
In this section we evaluate the effect of a constant elongation
$\kappa$ on the different terms in the gyrokinetic equation assuming
large aspect ratio $\epsilon=r/R_0\ll 1$ and unshifted flux surfaces
($R_0$ is independent of $r$).

Defining a coordinate basis $\{r, \theta,\zeta \}$ through $ x=R
\cos(-\zeta)$, $y= R \sin(-\zeta)$, $z= \kappa r \sin \theta$, with
$R= R_0 +r \cos\theta$, where $\{x,y,z\}=\rv$ is the Cartesian system,
we find that the Jacobian is
\begin{equation}
\mJ=\left|\frac{\p (x,y,z)}{\p (r,\theta,\zeta)}\right|=r\kappa(R_0 +r
\cos\theta)=r R\kappa.
\label{mj}
\end{equation}
We may also calculate the Jacobian of the coordinate basis $\{\psi,
\theta,\zeta\}$ where $r=r(\psi)$ with the poloidal magnetic flux
  $2\pi\psi$, to find
\begin{equation}
\mJ_\psi=\left(\p\psi/\p r\right)^{-1}\mJ\quad=\left(B\bv\cd\na\theta\right)^{-1},
\label{mjpsi}
\end{equation}
where the magnetic field $\Bv=B\bv$ with $B=|\Bv|$.  The gradient
vectors can be evaluated to be
%\begin{align}
% \na r=&\left(\frac{\p \rv}{\p \theta}\times\frac{\p \rv}{\p
%   \zeta}\right)\mJ^{-1}=\begin{Bmatrix} \cos\theta\cos\zeta
% \\ -\cos\theta\sin\zeta \\ \kappa^{-1}
% \sin\theta\end{Bmatrix},\nonumber \\
%  \na \theta=&\left(\frac{\p
%   \rv}{\p \zeta}\times\frac{\p \rv}{\p
%   r}\right)\mJ^{-1}=\frac{1}{r}\begin{Bmatrix} -\sin\theta\cos\zeta
%   \\ \sin\theta\sin\zeta \\ \kappa^{-1}
%   \cos\theta\end{Bmatrix},\\
%   \na \zeta=&\left(\frac{\p \rv}{\p
%     r}\times\frac{\p \rv}{\p
%     \theta}\right)\mJ^{-1}=\frac{1}{R_0+r\cos\theta}\begin{Bmatrix}
%     -\sin\zeta \\ -\cos\zeta \\ 0 \end{Bmatrix},\nonumber
%\label{grads}
%\end{align}
\begin{align}
 \na r=&\left(\frac{\p \rv}{\p \theta}\times\frac{\p \rv}{\p
   \zeta}\right)\mJ^{-1}=\left\{\cos\theta\cos\zeta
,\, -\cos\theta\sin\zeta ,\, \kappa^{-1}
 \sin\theta\right\},\nonumber \\
  \na \theta=&\left(\frac{\p
   \rv}{\p \zeta}\times\frac{\p \rv}{\p
   r}\right)\mJ^{-1}=\frac{1}{r}\left\{ -\sin\theta\cos\zeta
   ,\, \sin\theta\sin\zeta ,\, \kappa^{-1}
   \cos\theta\right\},\\
   \na \zeta=&\left(\frac{\p \rv}{\p
     r}\times\frac{\p \rv}{\p
     \theta}\right)\mJ^{-1}=\frac{1}{R_0+r\cos\theta}\left\{
     -\sin\zeta,\, -\cos\zeta,\, 0 \right\},\nonumber
\label{grads}
\end{align}
from which we find
\begin{equation}
%\begin{aligned}
%\begin{align}
\label{gradsquares}
%\label{eq:ntprofs}
\eqalign{
%\label{gradsquares}
 |\na r|^2=& 1-(1-1/\kappa^{2})\sin^2\theta, \\ %\nonumber, \\ 
 |\na \theta|^2=&
 \left[1+(\kappa^2-1)\sin^2\theta\right]/(r\kappa)^2,%\label{gradsquares}
 \\
|\na\zeta|^2=& 1/R^{2},\\%\nonumber\\
\na r\cd\na\theta=& \left[(1-\kappa^2)
  \cos\theta\sin\theta\right]/(r\kappa^2),%\nonumber,
%\end{aligned}
}
\end{equation}
%\end{align}
and $\na r\cd\na\zeta=0=\na \theta\cd\na\zeta$. The strength of the
poloidal magnetic field is given by
$B_p=|\na\psi|/R=(\p \psi/\p r) |\na
  r|/R=B_{p0}\left[1-(1/\kappa^2)\sin^2\theta\right]^{1/2}$, with
$B_{p0}=R_0^{-1}\p\psi/\p r$.

We introduce the binormal variable $\al$ through the following
representation of the magnetic field $\Bv=\naa\times\nap$. This
representation is consistent with the usual $\Bv=\naz\times\nap+I
\naz$ when $\al=\ze-\int_0^{\tht} d \tht I/(R^2\Bv\cd\nabt)\approx
\ze-q\tht$, where we define the safety factor as
$q=B_0r\kappa/(R_0B_{p0})$ and neglect $\epsilon$ corrections.

When the fluctuating quantities X have the form
$X=X(\theta)\exp\{-i\omega t+i\kps\psi-in\al\}$, where $\theta$ is the
ballooning angle, the perpendicular wave number defined by $\na_\perp
X=i\kpev X$ can be written as
\begin{equation}
\kpev=[\kps+n\tht(\p q/\p
  \psi)]\nap-n(\naz-q\nabt)=k_y[r\nabt+s(\theta-\theta_0)\nar],\label{kperpv}
\end{equation}
where we introduced the binormal wave number $k_y=nq/r$,
$\theta_0=-k_\psi R_0B_{p0}/k_y s$, and recalled the definitions
$s=(r/q)\p q/\p r$ and $R_0B_{p0}=\p \psi/\p r$, furthermore, in
$\kpev$ we neglected a term of the size $k_yr/(qR_0)$ as small in
$\epsilon$.

From Eq.~(\ref{kperpv}) we find the perpendicular wave number appearing in
finite Larmor radius terms is given by
%\begin{equation}
$\kpe=k_y\left[r^2|\nabt|^2+s^2(\theta-\theta_0)^2|\nar|^2+
  2rs(\theta-\theta_0)\nar\cd\nabt \right]^{1/2}$,
%\label{kperp}
%\end{equation}
where the gradient expressions are given by Eq.~(\ref{gradsquares}).

Then we turn to the drift terms appearing in Eq.~(\ref{gke}) assuming that
$\kpev$ is orthogonal to $\nar$ at the outboard mid-plane, that is,
$\theta_0=0$. First we consider the radial $\mathbf{\Ev\times\Bv}$
drifts
\begin{equation}
\omega_E=\left[\frac{1}{B^2}\Bv\times\nabt\frac{\p \phi_E}{\p
    \theta}\right]\cd\kpev=\frac{k_y}{B^2}\frac{\p \phi_E}{\p
  \theta}\frac{\p\psi}{\p r} \left[(\naa\times\nar)\times\nabt
  \right]\cd\left[ s\theta\nar+r\nabt\right],
\end{equation}
where we write $\Bv=(\p \psi/\p r)\naa\times\nar$, and used that
$\p\phi_E/\p r$ does not contribute to the radial drift and $(\p/\p
\alpha|_{\theta,r})\phi_E=\p\phi_E/\p \zeta=0$. Using the definition
of $\al$ we find
\begin{equation}
\omega_E=C\left\{\left[\left(\naz-q\nabt-\tht(\p q/\p r)\nar
  \right)\times \nar\right]\times\nabt \right\}\cd(s\tht\nar+r\nabt),
\label{ome2}
\end{equation}
with $C=(k_y/B^2)(\p \phi_E/\p \tht)(\p \psi/\p r)$. After some algebra
Eq.~(\ref{ome2}) reduces to
\begin{equation}
\omega_E=-Cqs\tht\left[ |\nar|^2|\nabt|^2-(\nar\cd\nabt)^2 \right].
\label{ome3}
\end{equation}
Using Eq.~(\ref{gradsquares}), the expression in the square bracket in
Eq.~(\ref{ome3}) can be simplified to $1/(r\kappa)^2$, which, together
with $\p\psi/\p r\approx Br\kappa/q$ yields the result, Eq.~(\ref{ome}).

Next we evaluate the magnetic drift frequency of species $a$ in the
$\beta\ll 1$ limit, when the magnetic drift velocity is given by
\begin{equation}
\vv_{Ma}=\frac{\Bv\times \na
  B}{B^2\Omega_a}\left(\frac{\vpe^2}{2}+\vpa^2\right)\approx-\frac{\bv\times\na
  R}{R\Omega_a}\left(\frac{\vpe^2}{2}+\vpa^2\right),
\label{vd}
\end{equation}
where $\na R=\cos\tht\nar-r\sin\tht\nabt$ and $\bv=B^{-1}(\p\psi/\p
r)\naa\times\nar$. The magnetic drift frequency is given by
\begin{equation}
\omega_{Da}=\kpev\cd\vv_{Ma}=-(R\Omega_a)^{-1}
\left(\vpe^2/2+\vpa^2\right)\kpe\cd\bv\times\na R.
\label{omd1}
\end{equation}
The triple product in Eq.~(\ref{omd1}) is
\begin{equation}
\kpe\cd\bv\times\na R=\frac{k_y}{B}\frac{\p \psi}{\p
  r}\left[(\naz-q\nabt)\times\nar\right]\times(\cos\tht
\nar-r\sin\tht\nabt)\cd(s\tht\nar+r\nabt),
\label{trip1}
\end{equation}
where we used the definition of $\alpha$. It can be easily shown that
the $\naz$ term in Eq.~(\ref{trip1}) is exactly zero due to toroidal
symmetry, and after dropping $\nar\times \nar$ and $\nabt\times\nabt$
terms Eq.~(\ref{trip1}) reduces to
\begin{equation}
\kpe\cd\bv\times\na R=-qr\frac{k_y}{B}\frac{\p
  \psi}{\p r}(\nabt\times\nar)\cd(\nabt\times\nar)(\cos\tht+s\tht\sin\tht).
\label{trip1mod}
\end{equation}
We may write $
(\nabt\times\nar)\cd(\nabt\times\nar)=-|\nabt|^2|\nar|^2-(\nabt\cd\nar)^2=-(r\kappa)^{-2}$,
as we found after Eq.~(\ref{ome3}). Combining this result with
Eqs.~(\ref{omd1}) and (\ref{trip1}), and recalling $\p\psi/\p r\approx
Br\kappa/q$ we arrive at the result
\begin{equation}
\omega_{Da}=-\frac{k_y}{\kappa R\Omega_a}
\left(\frac{\vpe^2}{2}+\vpa^2\right)(\cos\tht+s\tht\sin\tht).
\label{omdf}
\end{equation}

Finally we consider the diamagnetic frequency $\omega_{\ast a}^T$ that
is defined through the relation
\begin{equation}
i \omega_{\ast a}^T \frac{e_a
  f_{a0}}{T_a}\langle\phi\rangle_{\mathbf{R}}=-\langle\vv_{E}^t\rangle_{\mathbf{R}}\cd
\nar \frac{\p f_{a0}}{\p r},
\label{omsdef0}
\end{equation}
with $\langle\vv_{E}^t\rangle_{\mathbf{R}}$ the turbulent
$\Ev\times\Bv$ drift frequency gyro averaged keeping the guiding
center position fixed. Using
$\langle\vv_{E}^t\rangle_{\mathbf{R}}=i\bv\times\kpev\langle\phi\rangle_{\mathbf{R}}/B$
we can express $\omega_{\ast a}^T$ as
\begin{equation}
\omega_{\ast a}^T =-\frac{T_a}{e_a B}\bv\times\kpev\cd\nar\frac{\p \ln
  f_{a0}}{\p r}.
\label{omsdef1}
\end{equation}
Using the expression for $\bv$ after Eq.~(\ref{vd}), together with
Eq.~(\ref{kperpv}) and $\naa=\naz-(qs/r)\nar-q\nabt$ we write
Eq.~(\ref{omsdef1})
\begin{equation}
\omega_{\ast a}^T =-\frac{k_y T_a}{e_a B^2}\frac{\p\psi}{\p r}\frac{\p
  \ln f_{a0}}{\p r}\nar\cd\left[\left(\naz-\frac{qs}{r}\nar-q\nabt\right)
  \times\nar\right]\times (r\nabt+s\tht\nar).
\label{omsdef2}
\end{equation}
The $\nar\cd\dots$ part of the expression in Eq.~(\ref{omsdef2}) can be
rewritten as $-qr\left[ |\nar|^2|\nabt|^2-(\nar\cd\nabt)^2
  \right]=-qr/(r\kappa)^2$ which, together with $\p\psi/\p r\approx
Br\kappa/q$, leads to the result
\begin{equation}
\omega_{\ast a}^T =\frac{k_y T_a}{\kappa e_a B}\frac{\p \ln f_{a0}}{\p
  r}.
\label{omsfin}
\end{equation}
We found that all the drift frequencies appearing in Eq.~(\ref{gke}) scale
as $1/\kappa$.

To find that the parallel streaming term $\vpa\bv\cd\nabt (\p/\p\tht)$
in Eq.~(\ref{gke}) is unaffected by the elongation, $\bv\cd\nabt\approx
1/qR_0$, we recall the expression for $\mJ_\psi$ in Eqs.~(\ref{mjpsi}) and
(\ref{mj}), and the definition of the safety factor $q=B_0r\kappa/(R_0B_{p0})$.

\section*{Acknowledgments}
The simulations were performed on resources provided by the Swedish
National Infrastructure for Computing (SNIC) at PDC Center for High
Performance Computing (PDC-HPC).  The authors would like to thank
J.~Candy for providing the \gyro~code, S.~Moradi for fruitful
discussions and the entire C-Mod team for excellent maintenance and
operation of the tokamak. IP was supported by the Intenational Postdoc grant from Vetenskapsr{\aa}det.    

\bibliographystyle{unsrt}

\end{document}